\documentclass{article}

\usepackage[english]{babel}
\usepackage{algorithm}
\usepackage{bbm}
\usepackage{amsthm}

\usepackage[letterpaper,top=2cm,bottom=2cm,left=3cm,right=3cm,marginparwidth=1.75cm]{geometry}

\usepackage{amsmath}
\usepackage{graphicx}
\usepackage[colorlinks=true, allcolors=blue]{hyperref}
\usepackage{float}
\usepackage{natbib}

\usepackage[usenames,dvipsnames]{xcolor} 
\usepackage[ulem=normalem]{changes}
\usepackage{manyfoot, marginnote}
\usepackage{xifthen}

\newcommand{\Maricol}{WildStrawberry}

\newcommand{\Mari}{\color{\Maricol}}

\makeatletter
\newcommand\coluwave[2][\Mari]{\bgroup\markoverwith
{#1{\lower3.5\p@\hbox{\sixly \char58}}}\ULon{#2}}

\newcommand\colsout[2][\Mari]{\bgroup\markoverwith
{#1{\rule[.5ex]{2pt}{0.8pt}}}\ULon{#2}}

\makeatother

\newcounter{comments}
\newcommand{\com}[3]{\ifthenelse{\isempty{#3}}{}%
{\refstepcounter{comments}\Footnotemark{$^\text{#2#1\arabic{comments}}$}%
\Footnotetext{#2#1\arabic{comments}}{~\textsl{#2#3}}\marginnote{#2\scriptsize[#1\arabic{comments}]}}}

\title{The power of visualizing distributional differences: Formal graphical $n$-sample tests } 
\author{\vspace{0.3cm}Konstantinos Konstantinou$^1$,Tom\' a\v s Mrkvi\v cka$^2$ and Mari Myllym\" aki$^3$\\ \vspace{0.2cm} $^1$Chalmers University of Technology and University of Gothenburg\\ \vspace{0.2cm}$^2$University of South Bohemia, Faculty of Agricultural and Technology\\ $^3$Natural Resources Institute Finland (Luke)}

\begin{document}
\maketitle

\begin{abstract}
Classical tests are available for the two-sample test of correspondence of distribution functions. 
From these, the Kolmogorov-Smirnov test provides also the graphical interpretation of the test results, in different forms.
Here, we propose modifications of the Kolmogorov-Smirnov test with higher power. The proposed tests are based on the so-called global envelope test which allows for graphical interpretation, similarly as the Kolmogorov-Smirnov test.  The tests are based on rank statistics and are suitable also for the comparison of $n$ samples, with $n \geq 2$. We compare the alternatives for the two-sample case through an extensive simulation study and discuss their interpretation. Finally, we apply the tests to real data.
Specifically, we compare the height distributions between boys and girls at different ages, the sepal length distributions of different flower species, 
and distributions of standardized residuals from a time series model for different exchange courses using the proposed methodologies.


\end{abstract}

\section{Introduction}

In statistical theory, hypothesis testing has a central role. Often, statisticians need to address whether the distribution of a population follows a theoretical distribution. These types of tests are often referred to as goodness-of-fit tests in the literature.  For instance, the Shapiro-Wilk test \citep{shaphiro1965analysis} is a goodness-of-fit test that tests for Gaussianity of the underlying population. However, the test lacks a graphical illustration of the test result. Alternatively, various graphical goodness-of-fit procedures \citep{kolmogorov1933sulla,warton2022eco,aldor-noiman_power_2013} for testing the Gaussianity of the underlying population also provide graphical visualization of the test outcome.   The graphical nature of the tests enables statisticians to investigate how the data distribution deviates from the null distribution, which makes these tests extremely popular.

Goodness-of-fit tests can be extended to investigate whether distributions of two populations are different, and if they are, how they differ. Statistical tests for comparing the distributions of two samples can be constructed by considering test statistics defined as deviations between the empirical cumulative distribution functions of the two samples. Classical examples of such tests are the two-sample Kolmogorov-Smirnov (KS) test \citep{kolmogorov1933sulla}, Cram{\'e}r von Mises test \citep{cramer1928composition} and Anderson-Darling test \citep{anderson1954test}. Among these tests, only the KS test can provide a graphical interpretation of the test results, i.e., a graphical illustration showing the reason for rejecting the null hypothesis. On the other hand, the KS test has some disadvantages. First and foremost, it is an asymptotic test, and hence it requires "large" sample sizes to be valid. Second, it is designed only for comparing two continuous distributions which limits its applicability. Third, it is more sensitive around the median of the two distributions and gives low statistical power when distributional differences lie in the tails \citep[see e.g.][]{aldor-noiman_power_2013}. To deal with these issues, we propose graphical non-parametric tests based on data permutations. The proposed tests are valid for comparing $n\geq 2$ samples of any sizes and achieve higher power than the KS test.

Statistical tests for the distribution comparison of two samples can be constructed using Monte Carlo methods. Such tests are non-parametric, based on data permutations and evaluation of a test statistic expressing a distributional contrast. 
For instance, the R \citep{R2023} package twosamples \citep{twosamples} implements permutation tests with contrasts defined as deviations between the empirical cumulative distribution functions \citep{dowd2020new}. Among others, permutation tests based on the  Kolmogorov-Smirnov, Kuiper \citep{kuiper1960tests} and Wasserstein \citep{vaserstein1969markov} deviations are available. Permutation tests can also be constructed by considering deviation measures between the two probability density functions. A rigorous review of potential test statistics, i.e., deviation measures between two densities, is presented by \citet{cha2007comprehensive}, who categorized the measures into seven main families. The Minkowski $L_p$ family, the $L_1$ family, the intersection family, the inner product family, the fidelity family, the squared $L_2$ family, and the Shannon's entropy family. Finally, distances combining ideas from measures from the aforementioned families are also possible. Implementations of the aforementioned distances are available in the R package 
philentropy \citep{Hajk2018}. However, similar to the classical Cram{\'e}r von Mises, Monte Carlo tests that are based on these other measures than the Kolmogorov-Smirnov measure do not provide any graphical interpretation of the test results and therefore are not considered in our analyses.

In this paper, we propose graphical tests for comparing the distributions of $n$ samples, with $n\geq2$. Graphical tests, such as tests based on the Kolmogorov-Smirnov statistic, are extremely popular in the literature but are known to have low statistical power in certain scenarios. To this end, we propose graphical non-parametric tests based on
data permutations achieving higher power than the Kolmogorov-Smirnov test. 
The graphical nature of the tests is necessary as it allows the user to study the reason for rejecting the null hypothesis. Thus, the test not only answers the question of whether the $n\geq2$ distributions are different but also explains how they differ. The proposed tests construct a $100(1-\alpha)$\% acceptance region for the test statistic under the null hypothesis. If the empirical test statistic falls outside this region at any point of the discretization, the null hypothesis is rejected.  The reason of the rejection can then be identified by examining the specific values where the empirical statistic deviates from the envelope. As it is important to be able to explain the differences, we provide illustrations and a rigorous discussion on how the user should interpret the test results in different scenarios.

The proposed tests are non-parametric, permutation tests based on
the global envelope testing framework proposed by \cite{MyllymakiEtal2017}. Global envelope tests provide a multiple testing adjustment procedure for testing statistical hypotheses involving multivariate or functional test statistics. That is, given a $d$-dimensional discretization of a test statistic $T$ and a significance level $\alpha$, the multiple testing issue is solved by controlling the family-wise error rate. 
The tests are based on ranking the "extremeness" of the empirical test statistic among simulations produced under the null hypothesis,  using a ranking measure. 
Thus, a prerequisite is that simulations of data under the null model are available. Generally, only one sample is available from each distribution, and hence null data are simulated using data permutations. For this purpose, a simple permutation of the data between the $n$ samples is a valid procedure to obtain simulations under the null hypothesis of equal distributions. As the tests are based on non-parametric ranking of the empirical and simulated test statistics, they make no assumptions regarding the distributions of the test statistics. Therefore, the user is flexible in choosing any set of test statistics for testing.  The only assumption is that the test statistics must be exchangeable under the permutation strategy. The exchangeability assumption guarantees that the test achieves the exact significance level $\alpha$ \citep[see discussion in][]{MyllymakiEtal2017}, given that the test statistics can be strictly ordered. The functional measures that we consider are designed to eliminate the ties between the test statistics. On the other hand, ties between the test statistics may still appear in some cases. In this case, the test is still valid but conservative. All permutation tests proposed in this study are based on the simple permutation scheme that satisfies exchangeability under the null hypothesis of equal distributions \citep{lehmann1986testing}. For our analyses, the data are assumed to be realizations from independent and identically (iid) distributed random variables under the null hypothesis of equal distributions in the $n$ groups . Hence, exchangeability is satisfied as it is a weaker assumption \citep{heath1976finetti}.

Suggested tests are based on test statistics capturing different aspects of the distributions under study.
Examples of such test statistics include the empirical cumulative distribution functions, kernel estimated density functions of the distributions, pairwise differences between the empirical cumulative distribution functions of two samples, pairwise comparisons of quantiles of two samples, as well as combinations of the aforementioned statistics. The quantile regression can be used for our aim, too, if the categorical covariate distinguishing the $n$ distributions is tested for its significance for all quantiles simultaneously 
\citep{mrkvivcka2023global}. 
This leads to the test statistic, which can be called the quantile regression process. We conducted a simulation study to compare the statistical power between the two-sample versions of the proposed graphical tests in different scenarios. As the focus is solely on comparing graphical tests, the power of the tests was compared with the KS test. According to our results, the proposed tests outperformed the classical KS test in terms of power in all studied settings. A brief discussion regarding the graphical interpretation of the tests in each scenario is presented. Finally, the proposed tests are applied to three real datasets.

The rest of the article is organized as follows. In Section \ref{sec:two-sample_tests}, we briefly present the asymptotic and permutation KS tests for comparing the distributions of two samples and describe the global envelope testing framework. In Section \ref{sec:n-sample_tests}, we describe how the suggested global envelope tests can be extended for comparison of the distributions of $n$ samples, and in Section \ref{sec: sim_study} we investigate the performance of the proposed tests concerning statistical power and graphical interpretation. Since various departures from the null model are investigated, this section can be treated as a dictionary of the visualizations obtained by different test statistics of those departures. In Section \ref{sec:data}, we apply the proposed test to three real datasets. Our results are discussed in Section \ref{sec:discussion}. The implementation of the proposed tests is available in the R package GET \citep{MyllymakiMrkvicka2023}.

\section{Two-sample tests}
\label{sec:two-sample_tests}

Assume that $X_1,..., X_{m_1}\sim F_1$ and $Y_1,..., Y_{m_2}\sim F_2$, are two independently and identically distributed samples from two unknown distributions $F_1$ and $F_2$, and that we wish to test the correspondence of the two distribution functions, i.e., the hypothesis 
\begin{equation}
\label{eq: H_0}
    H_0: F_1 = F_2 \text{ vs. } \ H_1: F_1\neq F_2.
\end{equation}
The equality sign "=" in Equation \eqref{eq: H_0} denotes that the two distributions, $F_1$ and $F_2$, are equal almost everywhere. That is, $F_1$ and $F_2$ are the same on a set of probability measure 1, but might be different on a set of probability measure 0.
A known class of tests for this two-sample case is based on a distance metric between the empirical cumulative distribution functions (ECDFs) 
\[
\widehat{F}_1(x) = \frac{1}{m_1}\sum_{i=1}^{m_1} \mathbf{1}(X_i\leq x)\ \text{ and }  \widehat{F}_2(x)=\frac{1}{m_2}\sum_{i=1}^{m_2} \mathbf{1}(Y_i\leq x)
\]
with $\mathbf{1}(\cdot)$ denoting an indicator function.

\subsection{The asymptotic two-sample Kolmogorov-Smirnov test}\label{sec:KS_asymp}

The asymptotic two-sample Kolmogorov-Smirnov test is based on the $\lVert \cdot \rVert_\infty$ norm of the difference between the two ECDFs. That is, the KS test statistic is given by

\begin{equation}\label{eq: KSstat}
D_{m_1+m_2} =\sqrt{M}\sup_x |\widehat{F}_1(x)-\widehat{F}_2(x)|
\end{equation}
where $M = m_1m_2/(m_1+m_2)$. The KS test statistic $D_{m_1+m_2}$ is distribution free, that is the distribution of $D_m$ under $H_0$ is independent of $F_1$ and $F_2$. Further, the asymptotic distribution of the test statistic \eqref{eq: 
 KSstat} under $H_0$ was characterized by \cite{kolmogorov1933sulla} and is known as the Kolmogorov distribution. For large sample sizes $m_1$ and $m_2$, and significance level $\alpha$, the  $100(1-\alpha)$\% KS envelope for the difference $F_1(x)-F_2(x)$ is given by
\[
\left[-\frac{c(\alpha)}{\sqrt{M}},\frac{c(\alpha)}{\sqrt{M}}\right]
\]
where $c(\alpha)$ is the critical value for the chosen significance level $\alpha$. Values of $c(\alpha)$ are listed in tables for different $\alpha$ \citep[see][]{smirnov1948table}. For instance, the value of $c(\alpha)$ for $\alpha=0.05$ is 1.36. The null hypothesis of the test is rejected if $D_{m_1+m_2} > c(a)$. Equivalently, the test provides the following graphical interpretation: $H_0$ is rejected if there exists an $x$ such that the difference $\widehat{F}_1(x)-\widehat{F}_2(x)$ lies outside the constant KS envelope. 

Other visualizations of the KS test were presented by \cite{doksum1976plotting}. These include visualizations for test statistics obtained by transformations of the KS statistic or visualizations of distributional contrasts other than the difference of the ECDFs. For instance, let $F_1(x) = F_2(x+\Delta(x))$, where $\Delta(x) $ is the amount of horizontal shift at $x$ needed to bring the distribution of $Y$ up to the distribution of $X$. Then, instead of visualizing the test result for the difference $F_1(x)-F_2(x)$, it is possible to visualize it for the shift $\Delta(x)$. That is, a simultaneous $100(1-\alpha)$\% confidence band for $\Delta(x)$ based on the KS statistic $D_{m_1+m_2}$ is given by

\begin{equation}
\left[\widehat{F}_2^{-1}\left(\widehat{F}_1(x)-\frac{c(\alpha)}{\sqrt{M}}\right)-x,\widehat{F}_2^{-1}\left(\widehat{F}_1(x)+\frac{c(\alpha)}{\sqrt{M}}\right)-x\right]
\label{eq: KS_SHIFT}
\end{equation}
Similarly, a simultaneous $100(1-\alpha)$\% confidence band for $\Delta(x) + x$ is given by
\begin{equation}
\left[\widehat{F}_2^{-1}\left(\widehat{F}_1(x)-\frac{c(\alpha)}{\sqrt{M}}\right),\widehat{F}_2^{-1}\left(\widehat{F}_1(x)+\frac{c(\alpha)}{\sqrt{M}}\right)\right].
\label{eq: KS_QQ}
\end{equation}
The band in Equation \eqref{eq: KS_QQ} corresponds to a confidence band for the quantile-quantile (QQ) plot, i.e., a graphical method where the quantiles of the two samples are plotted against each other. Similarly, the band in Equation \eqref{eq: KS_SHIFT} corresponds to a confidence band for the shift plot, i.e., the detrended QQ plot. On the other hand, they represent confidence regions around the empirical statistic, rather than acceptance regions around the statistic under the null hypothesis. For instance, $H_0$ is rejected if there exists an $x$ such that the band \eqref{eq: KS_SHIFT} does not include zero. The graphical tests for the QQ plot are available in the R package extRemes \citep{gilleland2005extremes}.

Unfortunately, tests based on the KS statistic are asymptotic tests and suitable only for comparing two samples coming from continuous distributions. Moreover, the tests are the most sensitive to deviations close to the median of the distribution. Therefore, they have low power when the distributional differences between the two distributions lie in the tails \citep{anderson1954test,aldor-noiman_power_2013}.

\subsection{The permutation two-sample Kolmogorov-Smirnov test}
\label{sec: KS_perm}

Another version of the two-sample KS test is based on permutations, rather than asymptotics \citep{praestgaard1995permutation}. As the permutation-based two-sample KS test is a Monte Carlo test, it is also applicable when $m_1$ and $m_2$ are small, and for non-continuous distributions. Firstly, the empirical KS statistic $D_{m_1+m_2}^0$ (see Equation \eqref{eq: KSstat}) is computed using the data.
Then, the following permutation scheme is used to construct simulated statistics $D_{m_1+m_2}^*$ under the null model \eqref{eq: H_0}. Let \[\mathbf{Z} = (Z_1,\ldots, Z_{m_1+m_2})=(X_1,\ldots, X_{m_1}, Y_1,\ldots, Y_{m_2})\] be the combined vector of the two samples. Then,  $\mathbf{X}^*=(X^*_1,\ldots, X^*_{m_1})$ is obtained by independently and randomly sampling $m_1$ elements from $\mathbf{Z}$ without replacement. Similarly, let $\mathbf{Y}^*=(Y^*_1,\ldots,Y^*_{m_2})$ denote the elements of $\mathbf{Z}$ not included in $\mathbf{X}^*$. This procedure creates two samples $\mathbf{X}^*$ and $\mathbf{Y}^*$ under the null model. Therefore, the distribution of $D_{m_1+m_2}$ under $H_0$ can be obtained from $s$ permutations of the data, by computing $D_{m_1+m_2}^i$ from each of the permutations $i=1,\dots,s$. Finally, the Monte Carlo $p$-value of the test is obtained by ranking $D_{m_1+m_2}^0$ among all test statistics $D_{m_1+m_2}^0, D_{m_1+m_2}^1,\ldots, D_{m_1+m_2}^s$. Graphical interpretation of this test can be obtained by considering the $\alpha(s+1)$th most extreme $D_{m_1+m_2}$ as the critical $c(\alpha)$. 
This global envelope corresponds to the ones proposed by \cite{Ripley1981}, having constant width. This method was improved by the global envelope tests proposed by \cite{MyllymakiEtal2017}.

\subsection{Global envelope tests}\label{sec: global_envelope}
The global envelope tests introduced in \cite{MyllymakiEtal2017} are non-parametric tests for functional or multivariate statistics initially developed to solve the multiple testing problem in spatial statistics, but extended to various other applications since then \citep{MyllymakiMrkvicka2023}. 
Let $\mathbf{x}=(x_1,\ldots,x_d)$ be a discretization of the domain where the functional test statistic of interest $\mathbf{T}=(T(x_1),\ldots, T(x_d))$ is evaluated. Similarly to the permutation KS test, global envelope tests are Monte Carlo tests and therefore they require the simulation of $s$ test statistics under the null model. In this work, simulated data under the null model are obtained using the simple permutation scheme detailed in Section \ref{sec: KS_perm}. 

Now let the empirical test statistic be denoted by $\mathbf{T}_0 =(T_0(x_1),\ldots, T_0(x_d))= (T_{01},\ldots, T_{0d})$ and the simulated statistics under the null model be denoted by $\mathbf{T}_1,\ldots,\mathbf{T}_s$.  Initially, a ranking measure $E$ needs to be chosen. Examples of such measures are the extreme rank length (ERL) measure \citep{NarisettyNair2016, MyllymakiEtal2017}, the continuous rank measure \citep{Hahn2015} and the area measure \citep{MrkvivckaEtal2022}. Moreover, for $i=0,\ldots,s$, let $E_i$ denote the resulting measures of the test statistics $\mathbf{T}_i$. Further, let $\prec$ be an ordering with the following interpretation: $E_i \prec E_j$ if $\mathbf{T}_i$ is more extreme than $\mathbf{T}_j$ with respect to the measure $E$. Therefore given a significance level $\alpha$, we can identify the critical value $E_{(\alpha)}$, i.e., the largest $E_i$ such that \begin{equation}
\sum_{i=0}^s (E_i \preceq E_{(\alpha)})\leq \alpha(s+1)\label{eq: E_a}\end{equation}
and consequently the index set $I_{(\alpha)}$ of the vectors which are less or as extreme as $E_{(\alpha)}$. Then, a $100(1-\alpha)\%$ global envelope is the band given by two vectors
$\mathbf{T}_{\text{low}}^{(\alpha)} = (T_{\text{low}\ 1
} ^{(\alpha)}, \dots, T_{\text{low}\ d } ^{(\alpha)})$
and $\mathbf{T}_{\text{upp}}^{(\alpha)} = (T_{\text{upp}\ 1    
    }^{(\alpha)}, \dots, T_{\text{upp}\ d    
    }^{(\alpha)})$
with
\begin{equation}
    \label{eq: gl_envelope}
    T_{\text{low}\ k
    } ^{(\alpha)}  = \min_{i\in I_{(\alpha)}}T_{ik } \text{ and }\quad  T_{\text{upp}\ k    
    }^{(\alpha)}= \max_{i\in I_{(\alpha)}}T_{ik }\quad \text{ for } k=1,\ldots,d.
\end{equation}

Global envelope tests not only produce a Monte Carlo $p$-value (based on the rank of $E_0$ among all $E_i$) but also provide the following graphical interpretation: If $\mathbf{T}_0$ goes outside the envelope $(\mathbf{T}_{\text{low}}^{(\alpha)},\mathbf{T}_{\text{upp}}^{(\alpha)}$) 
at any $x_k$, the null hypothesis is rejected and the $p$-value is less than $\alpha$. Furthermore, as the test is based on ranks it does not make any assumptions on the distribution of the test statistic $\mathbf{T}$ and hence is valid for any test statistic. The only assumption of the test is the exchangeability of the test vectors $\mathbf{T}_0,\ldots,\mathbf{T}_s$ under the permutation strategy. It is important to note, that the simple permutation strategy employed here satisfies the exchangeability assumption under the null hypothesis of equal distributions and hence, under the assumption that test vectors $\mathbf{T}_i$, $i=0,\ldots, s$, can be strictly ordered, the test achieves the desired nominal level.

For the comparison of distributions, we are interested in test statistics, which both lead to relatively high power and to an intuitive interpretation of the test results. 
A basic test statistic resembles the test statistic of the KS test, but instead of summarizing the differences $\widehat{F}_1(x) - \widehat{F}_2(x)$ for all values of $x\in\{x_1,\dots,x_d\}$ to a single number through Equation \eqref{eq: KSstat}, the test statistic (vector) is defined to consist of the differences of the two distributions for $x_1, \dots, x_d$,
\begin{equation}\label{eq: Tdiff}
\mathbf{T}^{\text{diff}} = \left( \widehat{F}_1(x_1) - \widehat{F}_2(x_1), \dots, \widehat{F}_1(x_d) - \widehat{F}_2(x_d) \right).
\end{equation} 
Another alternative is to choose the test statistic corresponding to the interpretation of the QQ plot (see Equation \eqref{eq: KS_QQ}). That is 
\begin{equation}\label{eq:Tqq}
\mathbf{T}^{\text{qq}} = \left( \widehat{F}_2^{-1}(\widehat{F}_1(x_1)), \dots, \widehat{F}_2^{-1}(\widehat{F}_1(x_d)) \right).
\end{equation}
In Section \ref{sec: sim_study}, we compare two further alternatives consisting of the ECDFs or kernel estimates of the probability density functions of all groups (see Table \ref{table: tests}), using the one-step combining procedure explained in \citet[][Appendix B]{MyllymakiMrkvicka2023}.
That is,
\begin{equation}\label{eq:Tecdf}
\mathbf{T}^{\text{ecdf}} = \left( \left( \widehat{F}_1(x_1), \dots, \widehat{F}_1(x_d) \right), \left( \widehat{F}_2(x_1), \dots, \widehat{F}_2(x_d) \right) \right),
\end{equation}
and 

\begin{equation}\label{eq:Tden}
\mathbf{T}^{\text{den}} = \left(\left(\widehat{f}_{1}^{b_1}(x_1),\ldots,\widehat{f}_{1}^{b_1}(x_d)\right),\left(\widehat{f}_{2}^{b_2}(x_1),\ldots,\widehat{f}_{2}^{b_2}(x_d)\right)\right),
\end{equation}
where $\hat{f_l}$, $l=1,2$, are the kernel density estimates of the probability density functions with bandwidths $b_l$. 
These statistics generalize directly to comparisons of $n$ groups by adding the ECDF/density values of the further groups to the test vector \eqref{eq:Tecdf} (see Section \ref{sec:n-sample_tests}). A pseudocode of the proposed tests is presented in Algorithm \ref{alg: pseudocode}. 

\begin{algorithm}
\caption{Global permutation test for comparing distributions.}
\label{alg: pseudocode}
    \begin{enumerate}
    \item Compute the test vector $\mathbf{T}_0$ for the observed data.
    \item Simulate $s$ replicates of data under the null hypothesis by permuting the data.
    \item Compute the test vectors for the $s$ simulated data, and obtain $\mathbf{T}_1,\dots,\mathbf{T}_s$.
    \item Apply a global envelope test to $\mathbf{T}_0,\mathbf{T}_1,\dots,\mathbf{T}_s$.
    \end{enumerate}
\end{algorithm}

The global envelope test is exact, i.e., the prescribed significance level is achieved exactly, if the probability of test vector ties (i.e., the measure $E$ does not distinguish between the two test vectors) is equal to 0. See \citet{MyllymakiEtal2017} for details.  
 However, the probability of test vector ties for the test described in Algorithm \ref{alg: pseudocode} with one of the test vectors in Equations \eqref{eq: Tdiff}-\eqref{eq:Tden} can be greater than 0. Thus, the test is conservative up to the level equal to $\alpha + $ the probability of test vector ties for any distributions $F_1$, $F_2$, and any test vector $\mathbf{T}$. This probability is rather small for the measures  $E$ used for global envelope testing.
Furthermore, the tests are conservative for any number of discrete values $d$ chosen for the discretization of the test statistic $T$, for any number of samples $n$, and for any dimension of $x$. 
Note that also  pointwise ties among $\mathbf{T}_0, \mathbf{T}_1, \dots, \mathbf{T}_s$ are likely for small or large values of $x$ in the test described in Algorithm \ref{alg: pseudocode} for the above test statistics \eqref{eq: Tdiff}-\eqref{eq:Tden}. 
To eliminate the influence of such argument values on the test, 
\citet{MyllymakiEtal2017} used the pointwise mid-ranks when constructing the ranking measure. 
This means that the test will not reject the null hypothesis due to the behavior of $\mathbf{T}_0$ at a specific $x$ where $\mathbf{T}_0$ obtains the most extreme value together with other $\mathbf{T}_i$.
Such pointwise ties can influence the graphical interpretation of the global envelope test, and therefore, we may in a rare case see a significant output where $\mathbf{T}_0$ lies on the edge of the envelope and not outside of the envelope.

\subsection{Quantile regression}\label{sec: QR}

Quantile regression \citep{koenker1978regression} is a statistical model that allows the study of covariate effects on the conditional quantile distribution of the response given a set of covariates. That is, given a response variable $\mathbf{Y}=(Y_1,\ldots, Y_m)$ and set of covariates $\mathbf{X}=(\mathbf{X}_1,\ldots, \mathbf{X}_m)$, the $\tau$- quantile of the conditional distribution $Y_i\mid \mathbf{X}_i$ for any quantile $\tau\in[0,1]$ is modelled by 

\begin{equation}
    Q_{Y_i\mid\mathbf{X}_i}(\tau)=\inf\{y:F_{Y_i\mid\mathbf{X}_i}(y)\geq \tau\}=\mathbf{X}_i^T\boldsymbol{{\beta}}(\tau), \quad i=1,\ldots , m,
\end{equation}
where $F_{Y_i\mid \mathbf{X}_i}$ is the conditional cumulative distribution function of $Y_i$ given $\mathbf{X_i}$ and the coefficient $\boldsymbol{\beta}(\tau)$ gives the effect of $\mathbf{X}$ on the $\tau$-quantile of the conditional response distribution. Here $\mathbf{X}_i$ and $\boldsymbol{\beta}(\tau)$ are $p$-dimensional vectors.

Recently, we proposed a statistical framework for performing simultaneous (global) inference for the quantile regression process $\boldsymbol{\beta}(\tau)$ for any discrete set of quantiles $\mathbf{\tau}\in\mathcal{T}=\{\tau_1,\ldots, \tau_d\}$ even in the presence of nuisance covariates \citep{mrkvivcka2023global}. The framework is based on global permutation-based envelope tests \citep{MyllymakiEtal2017} and allows for testing the following null hypothesis

\begin{equation}
    H_0 : \beta_l(\tau) = 0\text{ for all }  \tau\in \mathcal{T} \quad\text{ vs. }\quad H_1: \exists \ \tau\in\mathcal{T} \text{ such that }\beta_l(\tau) \neq 0
\end{equation}
for any $l=1, \ldots, p$. In this paper, we consider the special case of testing the effect of one categorical covariate with $n$ levels and no nuisance covariates. In this case, each level of the categorical covariate assigns every observation to the corresponding sample index, allowing us to study the effects of the categorical variable on the conditional quantile distribution of the response, i.e.\ whether the distributions of the $n$ samples are the same. Further, as the aim of this paper is to identify the best graphical test for comparing $n$ distributions, we did not consider any nuisance covariates.  Therefore, we used a simple permutation scheme to simulate under the null model. On the other hand, the test can be extended to include nuisance covariates through the permutation strategies discussed in \cite{mrkvivcka2023global}.
The test statistic is
\begin{equation}\label{eq:Tqr}
 \mathbf{T}^{\text{qr}} = \left(\widehat{\beta}(\tau_1),\ldots,\widehat{\beta}(\tau_d)\right)  
\end{equation}
where $\tau_1,\ldots,\tau_d \in[0,1]$ are $d$ quantiles and $\widehat{\beta}(\tau_k)$ is the estimated effect the categorical covariate $\mathbf{X}$ on the $\tau_k$-quantile of the conditional response distribution.

\subsection{Combining two or more test statistics }\label{sec: combining}
 
Global envelope tests can be extended to the case with more than one test statistic. Assume that $G$ vectors of test statistics 
\begin{equation}
\label{eq: G vectors}
\mathbf{T}_i^j=(T_{i1}^j,\ldots, T_{id_j}^j),\qquad j=1,\ldots, G, \quad i=1,\ldots,s,\quad d_j\geq1
\end{equation}
are available and we are interested in performing a test of the hypothesis \eqref{eq: H_0} using all these $G$ statistics. 
To give equal importance for each test statistic even when their dimensions $d_j$ differ, we used the following two-step combining procedure \citep[][Appendix B]{MyllymakiMrkvicka2023} for joint testing:
\begin{enumerate}
     \item For each $j=1,\dots,G$, rank the $s$ statistics $\mathbf{T}_i^j$, $i=1,\dots,s$, using a ranking measure $M$, as described in Section \ref{sec: global_envelope}. Let $m_i^j$ denote the measure associated with $\mathbf{T}_i^j$. 
     \item  Rank the test vectors $\mathbf{T}_i^*=(m_i^1,\ldots,m_i^G)$, $i=1,\dots,s$, using the one-sided ERL measure. Let $E_i$ denote the resulting measure for each $\mathbf{T}_i^*$.
 \end{enumerate} 
Now let $\alpha$ be the significance level, $E_{(\alpha)}$ be the largest $E_i$ satisfying Equation \eqref{eq: E_a}, and the index set $I_{(\alpha)}$ consist of the vectors less or as extreme as $E_{(\alpha)}$. Then the combined $100(1-\alpha)\%$ global envelopes are 
\begin{equation}
    T_{\text{low} \ k}^{(\alpha),j}=\min_{i\in I_{(\alpha)}}T_{ik}^j,\text{ and }\quad T_{\text{upp} \ k}^{(\alpha),j}=\max_{i\in I_{(\alpha)}}T_{ik}^j\quad\text{ for } k=1,\ldots d_j,\quad j=1,\ldots,G.
\end{equation}
For a more detailed explanation of the two-step combining procedure, the reader is referred to \cite{MyllymakiMrkvicka2023} and references therein.

\section{Comparing the distributions of $n$ samples}
\label{sec:n-sample_tests}

The two-sample tests of comparison of distribution functions introduced earlier can be generalized for comparison of $n$ samples. 
Assume that $\mathbf{X}^l = (X^l_1,\ldots, X^l_{m_l})$ with $l=1,\ldots, n$, are independently and identically distributed samples from $n$ distributions $F_l$, and that we are interested in testing the null hypothesis 
\begin{equation}
    H_0: F_1 = F_2 =\ldots = F_n \ \text{ vs } \ H_1:\exists \ 
 i, j\in\{1,\ldots,n\} \text{ such that } F_i\neq F_j.
    \label{eq: H0_nsamples}
\end{equation}

The hypothesis \eqref{eq: H0_nsamples} can be tested using classical methods, for example employing multiple two-sample tests such as the KS test and then applying a multiple testing correction such as the Holm-Bonferroni correction \citep{holm1979simple}. Unfortunately, this procedure is known to be conservative for large number of comparisons and dependent tests \citep{abdi2010holm}, which results in a loss of statistical power. Further, this procedure does not provide a graphical interpretation of the test results.

The proposed global test for the ECDF/density values, e.g., the test vector of Equation \eqref{eq:Tecdf}, can be extended for the case of $n$ samples by considering test vectors of the form 
\begin{equation}\label{eq:nsampleT}
\mathbf{T}=\left(\mathbf{T}_1,\ldots, \mathbf{T}_n\right),
\end{equation} 
where
$\mathbf{T}_{l} = (T_l(x_i), \dots, T_l(x_d))$ is the $l$th test vector evaluated at $(x_1,\dots,x_d)$.
The test can indicate the values of $x$ and also the samples $l$ which are the reasons for the possible rejection of the null hypothesis. Alternative specifications of the test statistics, can also allow identification of which of the pairs of the samples differ from each other, in a similar manner as in functional analysis of variance and linear model setups used in \citet{MrkvickaETal2020, MrkvickaEtal2021b}. For instance, the test statistics of Equations \eqref{eq:Tqq} and \eqref{eq: Tdiff} can be straightforwardly extended to the case of three or more groups. 
That is, let
\begin{equation}\label{eq:Tqq_lk}
\mathbf{T}_{lk}^{\text{qq}} = \left( \widehat{F}_k^{-1}(\widehat{F}_l(x_1)), \dots, \widehat{F}_k^{-1}(\widehat{F}_l(x_d)) \right),
\end{equation}
and take
\begin{equation}
\label{eq: npairs}
\mathbf{T}^{\text{qq}}=\left( 
\mathbf{T}_{12}^{\text{qq}},
\mathbf{T}_{13}^{\text{qq}},
\dots,
\mathbf{T}_{(n-1)n}^{\text{qq}}
\right),
\end{equation} 
with the number of QQ-statistics equal to $n(n-1)/2$, i.e.\ the number of distinct pairs of samples. The above test statistics can also be used jointly by means of a combined global envelope test similar to that of Section \ref{sec: combining}. 

\section{Simulation study to compare different test statistics}\label{sec: sim_study}
We conducted a simulation study to investigate the performance of the proposed global envelope tests in terms of power under different scenarios. 
All global envelope tests achieve correct nominal levels when the test statistics can be strictly ordered and are exchangeable under the permutation strategy, and they have been explored in several simulation setups earlier. 
To ensure that the possible ties of the test statistics do not have any prominent effect on the significance level of the tests, we explored the empirical significance level for the case of normally distributed samples. 
The significance levels were fine (see Appendix \hyperref[fig: TypeIerror]{A}).
Then we compared the power of the global envelope tests with the power of the KS test.

All tests included in the simulation study are listed in Table \ref{table: tests}.
The test statistic $\mathbf{T}^{\text{qr}}$ was evaluated at 100 discrete values of $\tau$ = ($\tau_1,\ldots,\tau_{100}$).
The rest of the test statistics were evaluated at $x=(x_1,\ldots, x_{100})$.
The test statistic $\mathbf{T}^{\text{den}}$ requires the choice of the bandwidths $b_l$ for the kernel density estimation of the probability density functions $f_l$, $l=1,2$ (see Equation \eqref{eq:Tden}). 
In this study, we did not consider the problem of optimally choosing the bandwidth and rather used Silverman's rule of thumb \citep{silverman2018density}.
All global envelope tests examined in the simulation were based on 5000 permutations and the ERL measure.

\begin{table}[H]
\caption{Description of the tests examined in the simulation study. For the specifications of the test statistics, see the referred equations, which were used with $d=100$.}
\centering
\scalebox{0.9}{
\begin{tabular}{l| c}
\label{table: tests}
 Test description 
 & Abbreviation\\
 \hline
 Global envelope test with $\mathbf{T}^{\text{ecdf}}$ (see Equation \eqref{eq:Tecdf})
 &  ECDF\\
 Global envelope test with $\mathbf{T}^{\text{den}}$ (see Equation \eqref{eq:Tden})
 &  DEN\\
 Global envelope test with $\mathbf{T}^{\text{diff}}$ (see Equation \eqref{eq: Tdiff})
 &  DIFF \\
Global envelope test with $\mathbf{T}^{\text{qq}}$ (see Equation \eqref{eq:Tqq})
&  QQ\\
 Global envelope test with $\mathbf{T}^{\text{qr}}$ (see Equation \eqref{eq:Tqr})
 &  QR\\
 
Combined global envelope test with  $\mathbf{T}^{\text{qq}}$ and $\mathbf{T}^{\text{den}}$
&  $\text{C}2$\\
Combined global envelope test with $\mathbf{T}^{\text{qq}}$,  $\mathbf{T}^{\text{diff}}$, $\mathbf{T}^{\text{ecdf}}$ and $\mathbf{T}^{\text{den}}$ 
&  $\text{C}4$\\
Asymptotic Kolmogorov Smirnov test &KS\\
\end{tabular}
}

\end{table}

We considered five experiments with a different distributional difference between the underlying distributions in each of them. More specifically, we performed two-sample comparisons of distribution tests with differences in the mean, variance, and skewness of the underlying distributions. Also, we investigated the case where one of the samples came from mixtures of distributions. In all experiments, the performance of the tests was evaluated with regard to statistical power, based on 1000 independent samples of size $N$, with $N$ = 10, 50, 100, and 200.

In addition to the power comparison results, in each simulation experiment setup, we present a detailed discussion regarding the graphical interpretation of the global tests. 
For illustration purposes, we show in figures only the last 100 components of the $\mathbf{T}^{\text{ecdf}}$ and  $\mathbf{T}^{\text{den}}$ statistics, i.e., how the second group differs from the mean of the groups (in the case of two groups, the first group has the opposite deviance from the mean). 
In the QR test, we always considered the first sample as our reference category. The resulting plots were obtained with 5000 permutations and sample sizes of 1000.
This sample size is that large that all tests considered reject the null hypothesis in all setups: the empirical statistic goes outside the global envelope for some values of $\tau$ or $x$.
The sample size was chosen this large for these illustrations, to concentrate on how different test statistics show the different types of deviances of the empirical statistic from the distribution under the null hypothesis \eqref{eq: H_0}.

\subsection{Difference in the tails}
\label{Sec: tails}
\begin{figure}[H]
    \centering
    \includegraphics[scale=0.25]{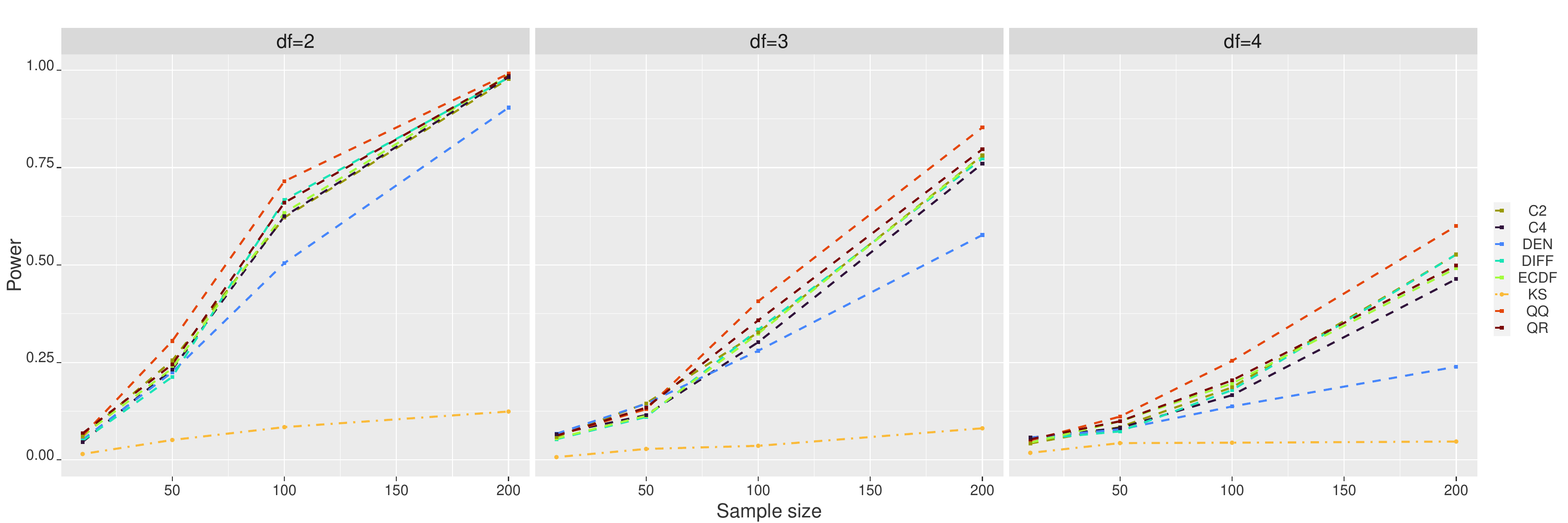}
    \caption{Power of the tests in Experiment \eqref{Ex:1} among 1000 simulated samples of different sizes (x-axis) for the different tests of Table \ref{table: tests} (colors). Each column represents the result for different degrees of freedom $df=2,3,4$ of the $t$ distribution.}
    \label{fig: tails}
\end{figure}

In Experiment \eqref{Ex:1}, we investigated the performance of the tests, in the case where the two distributions differ in the tails. For this purpose, we considered the following simulation setup:
\[
\left\{
\begin{array}{ll}\tag{I}
\label{Ex:1}
    X_i \sim \mathcal{N}(0,1)& \text{for } i=1,\ldots, N \\
    Y_i \sim t_{df}& \text{for } i=1,\ldots, N
\end{array}
\right.
\]
where $df=2,3,4$ are the degrees of freedom of the student $t$-distribution. In such a case, it is well established that the asymptotic KS test is inefficient in detecting distribution differences in the tails, which translates to low statistical power \citep{anderson1954test}. Indeed, as seen in Figure \ref{fig: tails}, the KS test has extremely low power compared to the proposed global envelope tests. Further, the power of the tests is negatively correlated with the number of degrees of freedom of the $ t$ distribution. This is because decreasing the number of degrees of freedom increases the deviation between the underlying distributions. Regarding the global envelope tests, the QQ test had the highest power while the DEN test had the lowest power.

\begin{figure}[H]
    \centering    \includegraphics[scale=0.25]{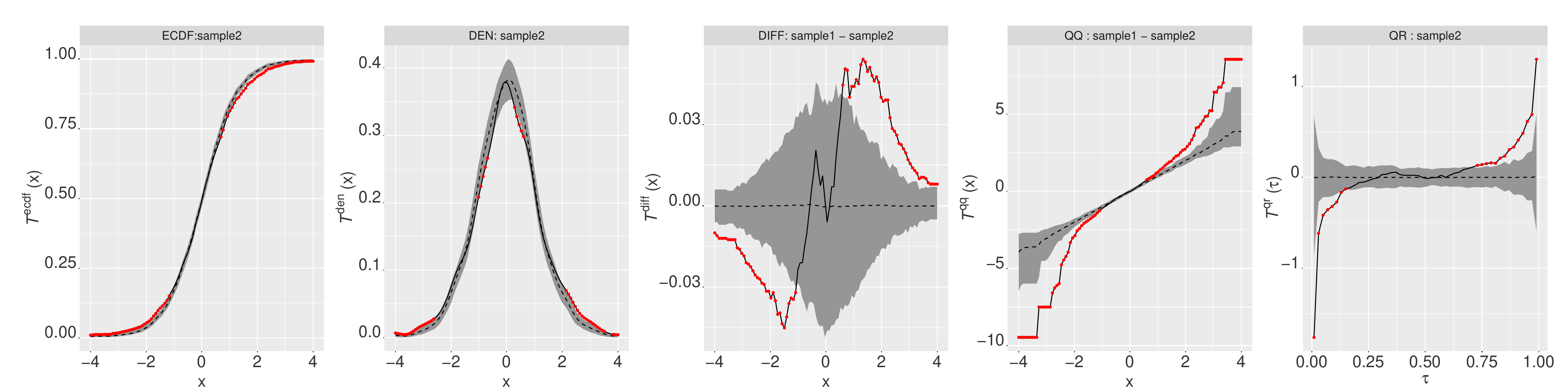}
    \caption{ Graphical interpretation of the global tests in Experiment \eqref{Ex:1} with $N=1000$. The solid curve represents the empirical test statistic $\mathbf{T}_0$ while the dashed line shows the expected values of the test statistic under the null hypothesis (see Equation \eqref{eq: H_0}). The shaded areas are the $95\%$ global envelopes constructed from 5000 permutations using the ERL measure. The points where $\mathbf{T}_0$ goes outside the global envelope are shown in red color. Each column shows the result of the global tests for the test statistic indicated in the titles (see Table \ref{table: tests}). For the ECDF and DEN tests, only the second element of the test statistic is shown. }
    \label{fig: graphical_tails}
\end{figure}

The graphical interpretation of the global tests for $df=3$ and different test statistics are shown in Figure \ref{fig: graphical_tails}. The corresponding results for $df$ = 2 and $df$ = 4 are similar and hence are not shown. 
The QR test rejects the null hypothesis due to differences between the two distributions in extreme quantiles indicating distributional differences in the tails for $\tau<0.2$ and $\tau>0.7$. In this test, the first sample, i.e., the sample from the normal distribution is used as a reference category, and hence the empirical statistic corresponds to the coefficient of the second sample, i.e., the student-$t$ distributed sample. The test indicates that the smallest values in the second sample are significantly smaller than the expected values under the reference category, while the largest values in the second sample are larger than the expected values under the reference category. The QQ test rejects the null hypothesis for $\mid x\mid>1$. The small quantiles ($x<-1$) of the $t$-distributed sample are significantly smaller than the quantiles expected under the null hypothesis \eqref{eq: H_0}. Similarly, the large quantiles ($x>1$) of the $t$-distributed sample are significantly larger than the quantiles expected under the null hypothesis. These observations indicate that the distribution of the second sample has heavier tails than the distribution of the first sample. 

The DIFF test also rejects the null hypothesis for $\mid x\mid>1$. Here, the test statistic $\mathbf{T}^{\text{diff}}$ is negative, i.e., $\widehat{F}_2(x)>\widehat{F}_1(x)$ for $x<-1$, and positive, i.e., $\widehat{F}_2(x)<\widehat{F}_1(x)$ for $x>1$. The same observation can be established from the ECDF test, where the second element of $\mathbf{T}^{\text{ecdf}}$ is shown. Thus, both test outputs indicate that the proportion of data with small values ($x<-1$) in the second sample is larger than the proportion of small values under the null hypothesis and there are more data in the second sample with large values ($x>1$) than what is expected under the null hypothesis. 
The DEN test 
shows that the second sample has higher density for $\mid x\mid>2$, as well as lower density for some $\mid x\mid<1$. These observations indicate that the distribution of the second sample has heavier tails than what is expected under the null hypothesis.

\subsection{Mean shift}
\label{Sec: mean_shift}
\begin{figure}[H]
    \centering
    \includegraphics[scale=0.4]{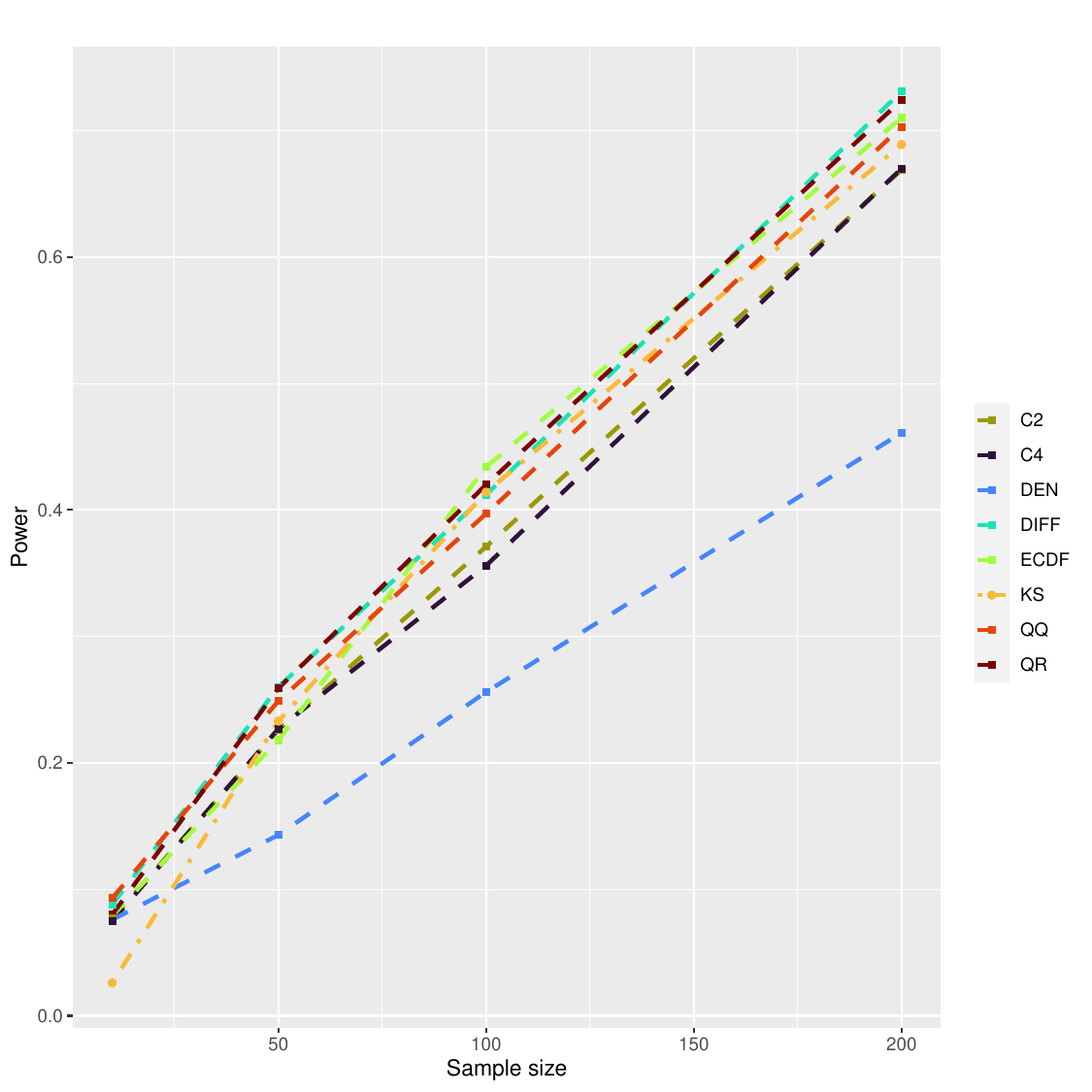}
    \caption{Power of the tests in Experiment \eqref{Ex:2} among 1000 simulated samples of different sizes (x-axis) for the different tests of Table \ref{table: tests} (colors).}
    \label{fig: mean}
\end{figure}

In Experiment \eqref{Ex:2}, we studied the performance of the tests, when the two distributions differ only in the mean. In this case, the two samples are obtained as follows

\[
\left\{
\begin{array}{ll}\tag{II}
\label{Ex:2}
    X_i \sim \mathcal{N}(0,1)& \text{for } i=1,\ldots, N \\
    Y_i \sim \mathcal{N}(0.3,1)& \text{for } i=1,\ldots, N
\end{array}
\right.
\]
In such a case, it is well known that the asymptotic KS test is powerful. On the other hand, the KS test was outperformed by the majority of the global tests (Figure \ref{fig: mean}). Overall, the QR test was the most powerful. As in Experiment \eqref{Ex:1}, the DEN test had the lowest power.

\begin{figure}[H]
    \centering   
    \includegraphics[scale=0.25]{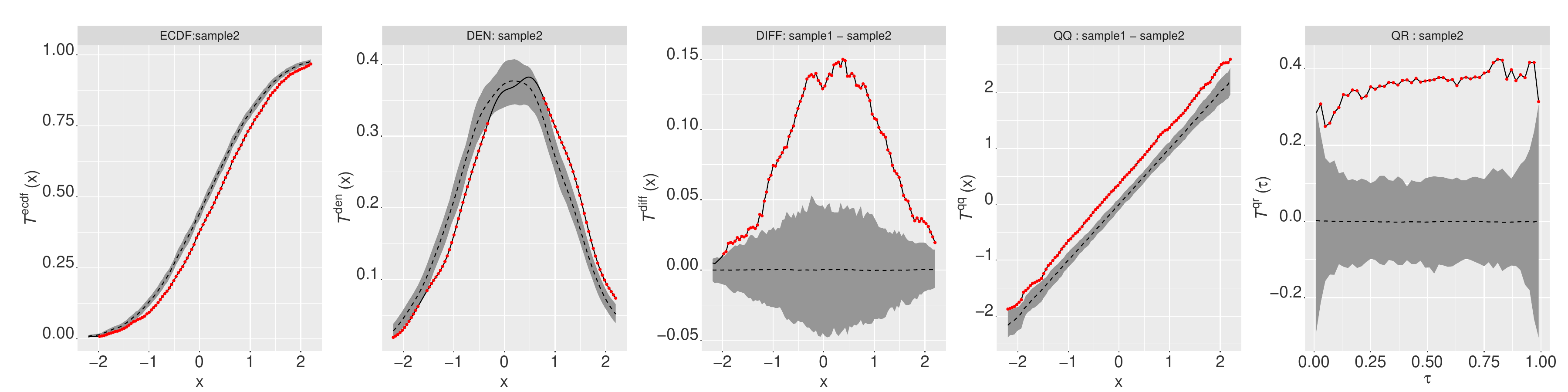}
    \caption{ Graphical interpretation of the global tests in Experiment \eqref{Ex:2} with $N=1000$. The solid curve represents the empirical test statistic $\mathbf{T}_0$ while the dashed line shows the expected values of the test statistic under the null hypothesis (see Equation \eqref{eq: H_0}). The shaded areas are the $95\%$ global envelopes constructed from 5000 permutations using the ERL measure. The points where $\mathbf{T}_0$ goes outside the global envelope are shown in red color. Each column shows the result of the global tests for the test statistic indicated in the titles (see Table \ref{table: tests}). For the ECDF and DEN tests, only the second element of the test statistic is shown.}
    \label{fig: graphical_mean}
\end{figure}

The graphical interpretation of the tests for different test statistics is displayed in Figure \ref{fig: graphical_mean}. 
Due to large sample size, all tests clearly reject the null hypotheses (the empirical test statistic is outside the 95\% global envelope for any value of $x$ or $\tau$), but with different reasoning. 
In the QR test we observe a significant (almost) constant positive quantile effect, i.e., $\widehat{\beta}(\tau)=0.3$ for most of the $\tau$ considered. Similarly, the QQ test rejects the null hypothesis for all $x$ considered. The empirical statistic is shifted uniformly from the line $y=x$, indicating that all values of the second sample are larger than the expected values under the null hypothesis. These observations, indicate that the two distributions differ by a location shift.

The DIFF test also rejects the null hypothesis for all $x$ considered. The test statistic is positive for all $x$ and therefore $\widehat{F}_2(x)<\widehat{F}_1(x)$ for all $x$. The same conclusion is derived from investigating the results of the ECDF test. 
According to the Figure \ref{fig: graphical_mean}, the ECDF of the second sample is shifted to the right compared to the mean ECDF of the two samples under the null hypothesis. This indicates that for a given $x\in[-2,2]$ we observe fewer data points smaller than $x$ than what is expected to be observed if the null hypothesis is true.  Similarly, the DEN test  
shows that the second sample has a density that is shifted to the right compared to the expected mean density under the null hypothesis. 

\subsection{Variance shift}
\begin{figure}[H]
    \centering
    \includegraphics[scale=0.4]{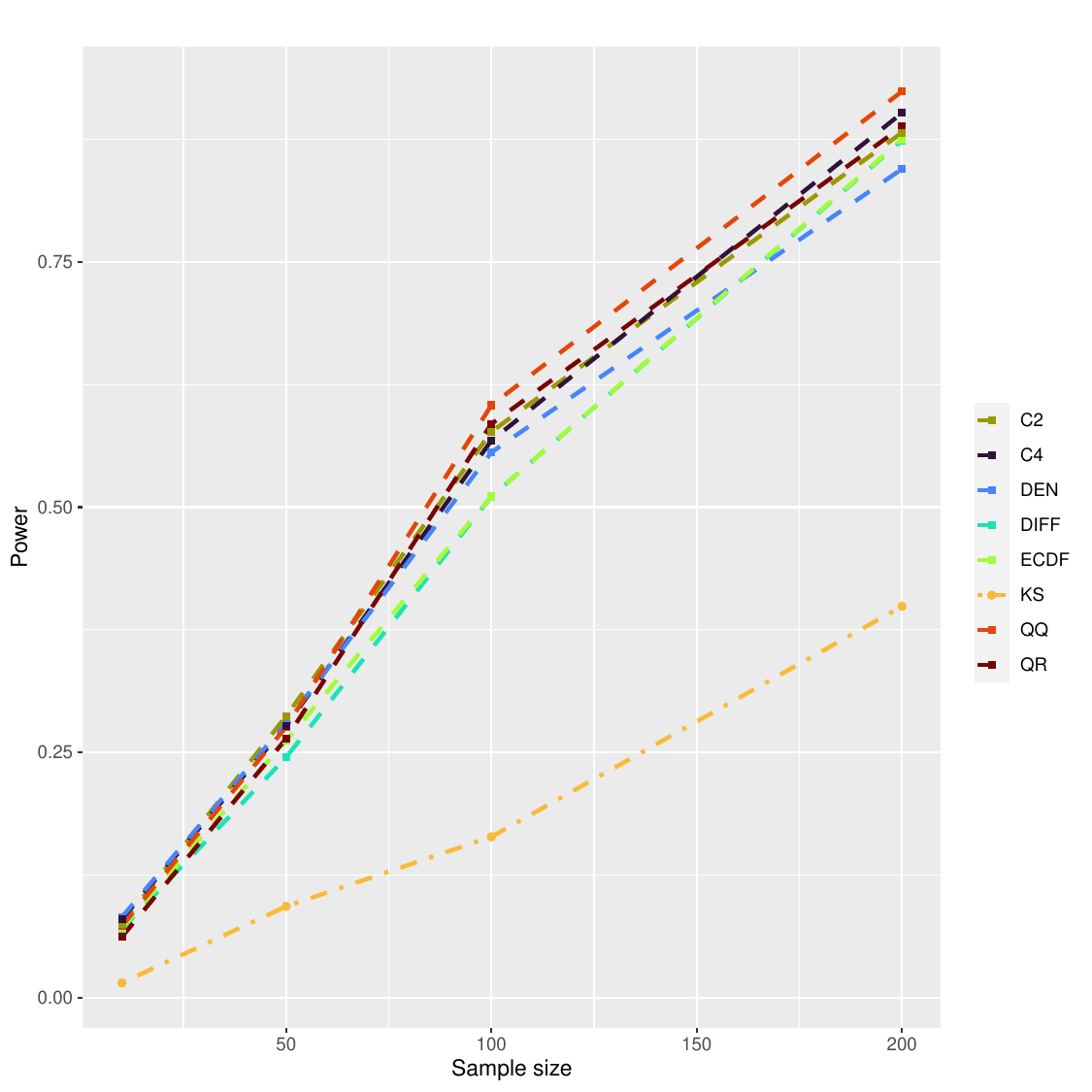}
    \caption{ Power of the tests in Experiment \eqref{Ex:3} among 1000 simulated samples of different sizes (x-axis) for the different tests of Table \ref{table: tests} (colors).}
    \label{fig: variance}
\end{figure}

In Experiment \eqref{Ex:3}, we studied the performance of the tests, when the two distributions differ only in the variance. In this case, the two samples are obtained as follows
\[
\left\{
\begin{array}{ll}\tag{III}
\label{Ex:3}
    X_i \sim \mathcal{N}(0,1)& \text{for } i=1,\ldots, N \\
    Y_i \sim \mathcal{N}(0,1.3)& \text{for } i=1,\ldots, N
\end{array}
\right.
\]
As seen in Figure \ref{fig: variance}, the asymptotic KS test had lower power than the proposed global envelope tests. The QQ test had the highest power while the ECDF test had the lowest power.  Overall, the differences between the global tests with different test statistics were rather small.

\begin{figure}[H]
    \centering    \includegraphics[scale=0.25]{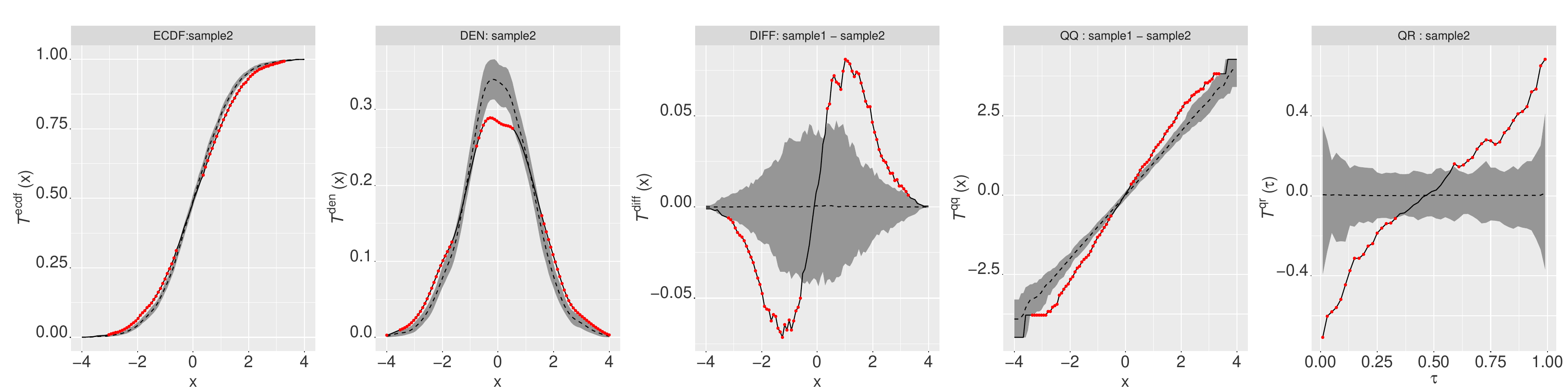}
    \caption{ Graphical interpretation of the global tests in Experiment \eqref{Ex:3} with $N=1000$. The solid curve represents the empirical test statistic $\mathbf{T}_0$ while the dashed line shows the expected values of the test statistic under the null hypothesis (see Equation \eqref{eq: H_0}). The shaded areas are the $95\%$ global envelopes constructed from 5000 permutations using the ERL measure. The points where $\mathbf{T}_0$ goes outside the global envelope are shown in red color. Each column shows the result of the global tests for the test statistic indicated in the titles (see Table \ref{table: tests}). For the ECDF and DEN tests, only the second element of the test statistic is shown. }
    \label{fig: graphical_var}
\end{figure}

The graphical interpretation of the tests for different test statistics is displayed in Figure \ref{fig: graphical_var}. The graphical interpretation of the variance shift is rather similar to the difference in the tails case, except for the fact that the variance shift shows the departures from the null hypothesis on longer domains.

\subsection{Mixture of normals}
\begin{figure}[H]
    \centering
    \includegraphics[scale=0.4]{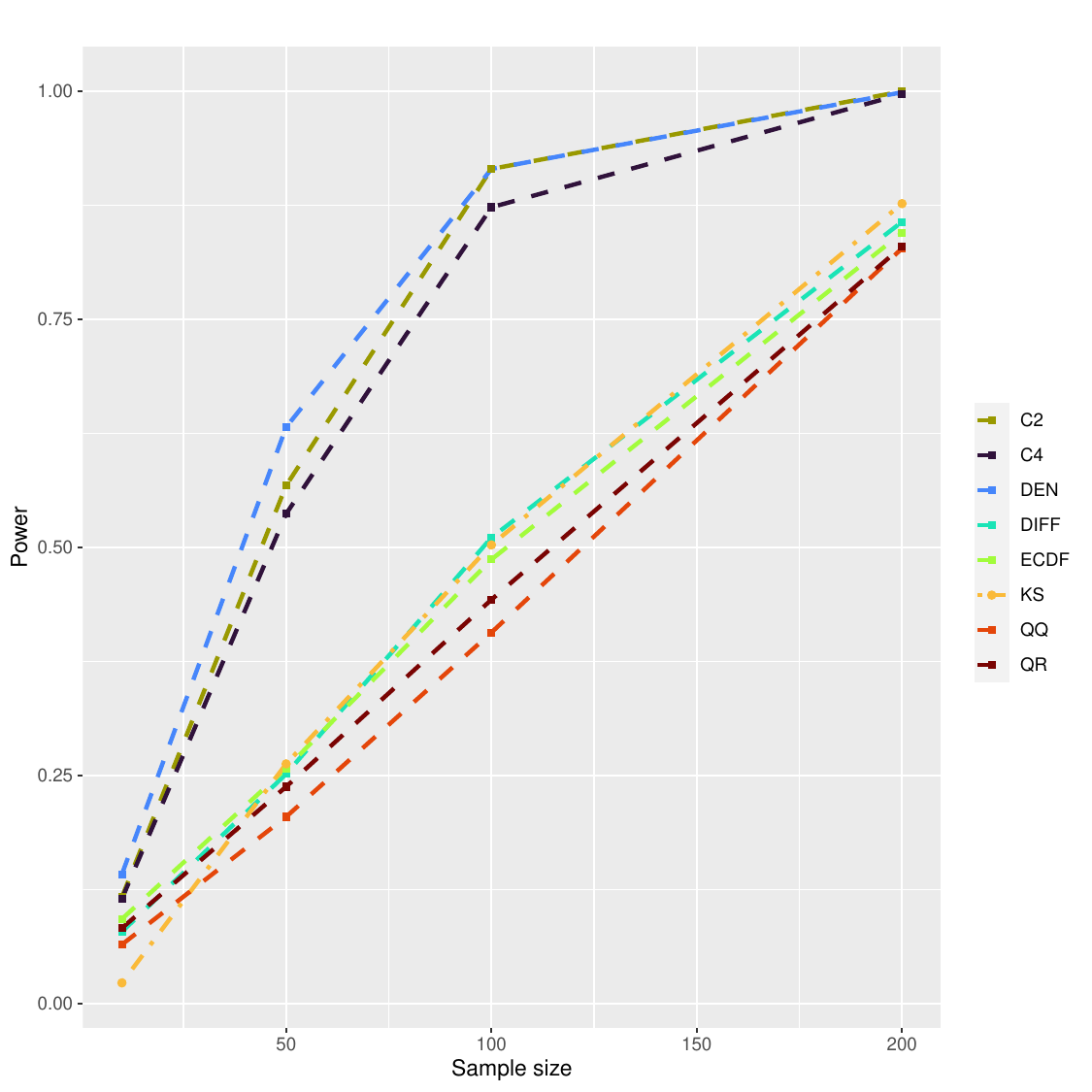}
    \caption{Power of the tests in Experiment \eqref{Ex:4} among 1000 simulated samples of different sizes (x-axis) for the different tests of Table \ref{table: tests} (colors).}
    \label{fig: mixture}
\end{figure}

In Experiment \eqref{Ex:4}, we studied the performance of the tests when one of the distributions is given as a mixture of two normal distributions and the other is the standard normal distribution. In this case, the two samples are obtained as follows
\[
\left\{
\begin{array}{ll}\tag{IV}
\label{Ex:4}
    X_i \sim \mathcal{N}(0,1) & \text{for } i=1,\ldots, N \\
    Z_i \sim \text{Bernoulli(0.5)}&\text{for } i=1,\ldots, N\\
    Y_i = Z_i\cdot \mathcal{N}(1,0.5) + ( 1 -Z_i)\cdot \mathcal{N}(-1,0.5)& \text{for }i=1,\ldots, N
\end{array}
\right.
\]
Unlike the results in the earlier sections, the DEN test was the most powerful test (see Figure \ref{fig: mixture}). However, it is worth mentioning that the combined global envelope tests that consider the $\mathbf{T}^{\text{den}}$ statistic have similar performance. On the other hand, the QQ test had the lowest statistical power.

\begin{figure}[H]
    \centering    \includegraphics[scale=0.32]{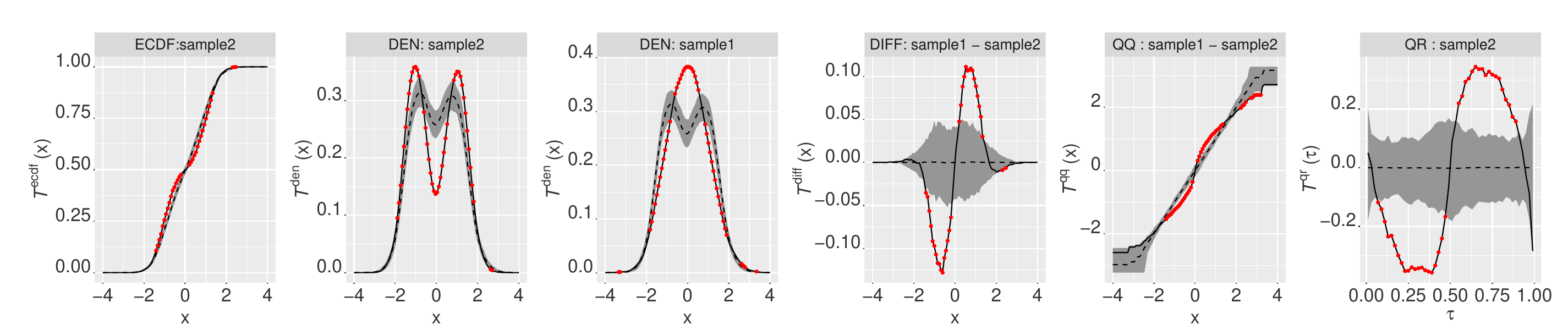}
    \caption{
   Graphical interpretation of the global tests in Experiment \eqref{Ex:4} with $N=1000$. The solid curve represents the empirical test statistic $\mathbf{T}_0$ while the dashed line shows the expected values of the test statistic under the null hypothesis (see Equation \eqref{eq: H_0}). The shaded areas are the $95\%$ global envelopes constructed from 5000 permutations using the ERL measure. The points where $\mathbf{T}_0$ goes outside the global envelope are shown in red color. Each column shows the result of the global tests for the test statistic indicated in the titles (see Table \ref{table: tests}). The second and third column together correspond to the two elements of the DEN test. For the ECDF test, only the second element of the test statistic is shown.}
    \label{fig: graphical_mixture}
\end{figure}

The graphical interpretation of the tests for different test statistics is displayed in Figure \ref{fig: graphical_mixture}. In the QR test,  we observe an S-shaped empirical statistic. This indicates that the quantile effect of the group covariate varies across the whole distribution. In particular, there is a significant negative quantile effect for $\tau\in[0.1,0.4]$, indicating that the $\tau$-quantiles of the second sample for $\tau\in[0.1,0.4]$ are smaller than the respective $\tau$-quantiles of the first sample. Further, there is a significant positive quantile effect for $\tau\in[0.6,0.9]$, indicating that the $\tau$-quantiles of the second sample for $\tau\in[0.6,0.9]$ are larger than the respective $\tau$-quantiles of the first sample. The graphical interpretation of the QQ test shows an S-shaped pattern as well. The second sample's quantiles $x\in[-1,0]$ are smaller and the second sample's quantiles $x\in[0,1]$  are larger than the quantiles expected under the null hypothesis. 

For the DIFF test, we observe an S-shape pattern indicating that   $\widehat{F}_2(x)>\widehat{F}_1(x)$ for $x\in[-1.5,0)$, and $\widehat{F}_2(x)<\widehat{F}_1(x)$ for $x\in(0,1.5]$.  
The results are also similar for the ECDF test. Finally, the DEN test provides a straightforward interpretation, that is the distribution of the second sample is a bimodal distribution (second plot of Figure \ref{fig: graphical_mixture}) and the distribution of the first sample is symmetric around 0 (third plot), and both of them are completely different than the mean behavior of the two samples under the null hypothesis of equal distributions. 

\subsection{Skewness}

\begin{figure}[H]
    \centering
    \includegraphics[scale=0.4]{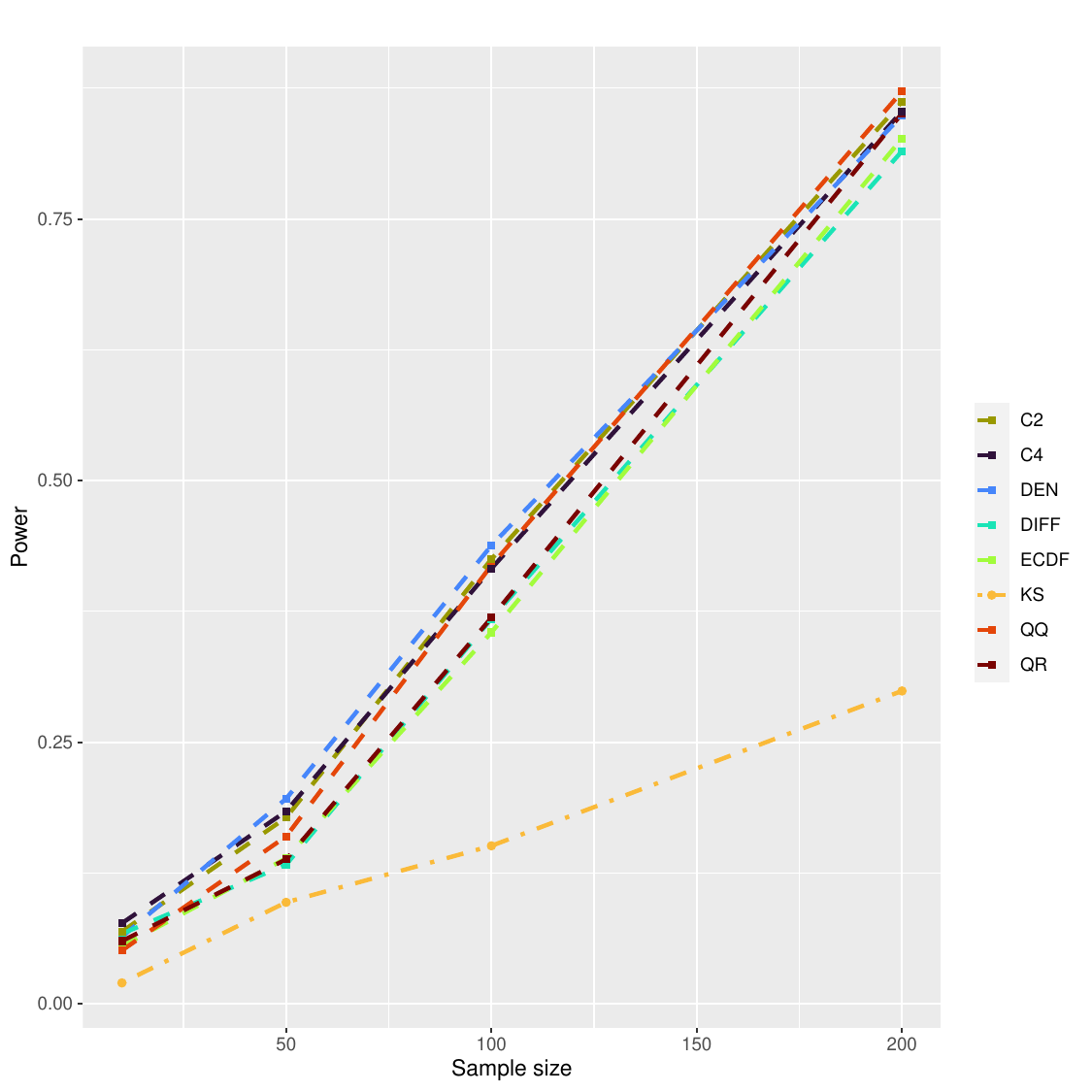}
    \caption{Power of the tests in Experiment \eqref{Ex:5} among 1000 simulated samples of different sizes (x-axis) for the different tests (colors).}
    \label{fig: skeweness}
\end{figure}

In Experiment \eqref{Ex:5}, we studied the performance of the tests when comparing a symmetrical distribution with a right-skewed distribution. In this case, the two samples were obtained as follows
\[
\left\{
\begin{array}{ll}\tag{V}
\label{Ex:5}
    X_i \sim \mathcal{N}(8,3.34)& \text{for } i=1,\ldots, N \\
    Y_i \sim \text{log-normal(2,0.4)}& \text{for } i=1,\ldots, N
\end{array}
\right.
\]
As seen in Figure \ref{fig: skeweness}, the proposed global envelope tests outperformed the asymptotic KS test in terms of power. The DEN test was the most powerful test for $N\leq100$, while the QQ test was the most powerful for $N=200$. Overall, the combined global envelope tests performed quite well, even though they were outperformed by the aforementioned tests. 

The QR test rejects the null hypothesis 
for extreme quantiles $\tau<0.1$ and $\tau >0.9$ as well as for intermediate quantiles $\tau\in[0.35,0.75]$. There is a negative effect at the extreme quantiles indicating that the smallest and largest values of the log-normal sample are smaller than under the reference category of the quantile model, i.e., the first sample stemming from the normal distribution.  Moreover, the quantile effect for $\tau\in[0.35,0.75]$ is positive indicating that the $\tau$-quantiles of the log-normal sample are larger than the respective $\tau$-quantiles under the reference category. Similarly, the QQ test rejects the null hypothesis for extreme quantiles $x<5$ and $x >15$ as well as for intermediate quantiles $x\in[7,10]$. 

\begin{figure}[H]
    \centering    \includegraphics[scale=0.32]{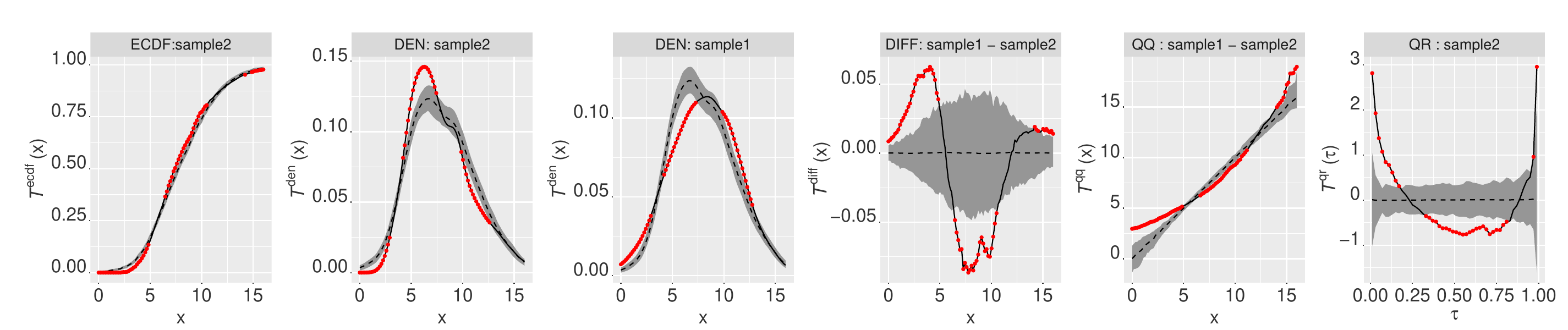}
    \caption{Graphical interpretation of the global tests in Experiment \eqref{Ex:5} with $N=1000$. The solid curve represents the empirical test statistic $\mathbf{T}_0$ while the dashed line shows the expected values of the test statistic under the null hypothesis (see Equation \eqref{eq: H_0}). The shaded areas are the $95\%$ global envelopes constructed from 5000 permutations using the ERL measure. The points where $\mathbf{T}_0$ goes outside the global envelope are shown in red color. Each column shows the result of the global tests for the test statistic indicated in the titles (see Table \ref{table: tests}). The second and third column together correspond to the two elements of the DEN test. For the ECDF, only the second element of the test statistic is shown.  }
    \label{fig: graphical_skew}
\end{figure}

The DIFF test, indicates that $\widehat{F}_2(x)>\widehat{F}_1(x)$ for $x\in[0,5]\cup[13,16]$ and that $\widehat{F}_2(x)<\widehat{F}_1(x)$ for $x\in[7,10]$. The same conclusion can be established from the ECDF plot. The former result, suggests that the proportion of data with small ($x<5$) and large ($x>13$) values in the second sample is larger than the proportion expected under the null hypothesis. The latter suggests that in the second sample, there is a larger proportion of values in the interval [7,10] than what is expected under the null hypothesis. Last but not least, the DEN test shows that the distribution of the first sample is symmetric (third plot of Figure \ref{fig: graphical_skew}) while the distribution of the second sample is a right skewed distribution (second plot), and that both distributions are different from the distribution under the null hypothesis of equal distributions.

\subsection{Recommendations}

The results from the simulation study indicated that the DEN test was slightly worse than the other tests in Experiments \eqref{Ex:1} and \eqref{Ex:2}. On the other hand, in Experiment \eqref{Ex:4} the DEN, C2, and C4 tests were the most powerful. Finally, all proposed tests achieved similar performance in Experiments \eqref{Ex:3} and \eqref{Ex:5}. The combined tests performed quite well in all examined cases, even though they were slightly outperformed by the best single test statistic-based tests in each scenario. As we typically do not know apriori how the distributions differ, we suggest using a combined test. Especially, we suggest using the $\text{C}2$ test as in most cases either the DEN or the QQ test was the most powerful. Moreover, this test provides a helpful graphical interpretation in each case as differences in the tails, location, or variances are well captured by the QQ test, and differences in the number of modes and location of the mode (skewness) are well captured by the DEN test. However, for $n$-sample comparisons with large $n$, we recommend using either the DEN, the QR, or the ECDF tests as their graphical interpretation includes only $n$ plots, while for the other tests the number of plots scale with $\mathcal{O}(n^2)$.

\section{Data Examples}\label{sec:data}
\subsection{Berkeley growth data}

\begin{figure}[H]
    \centering
    \includegraphics[scale=0.4]
{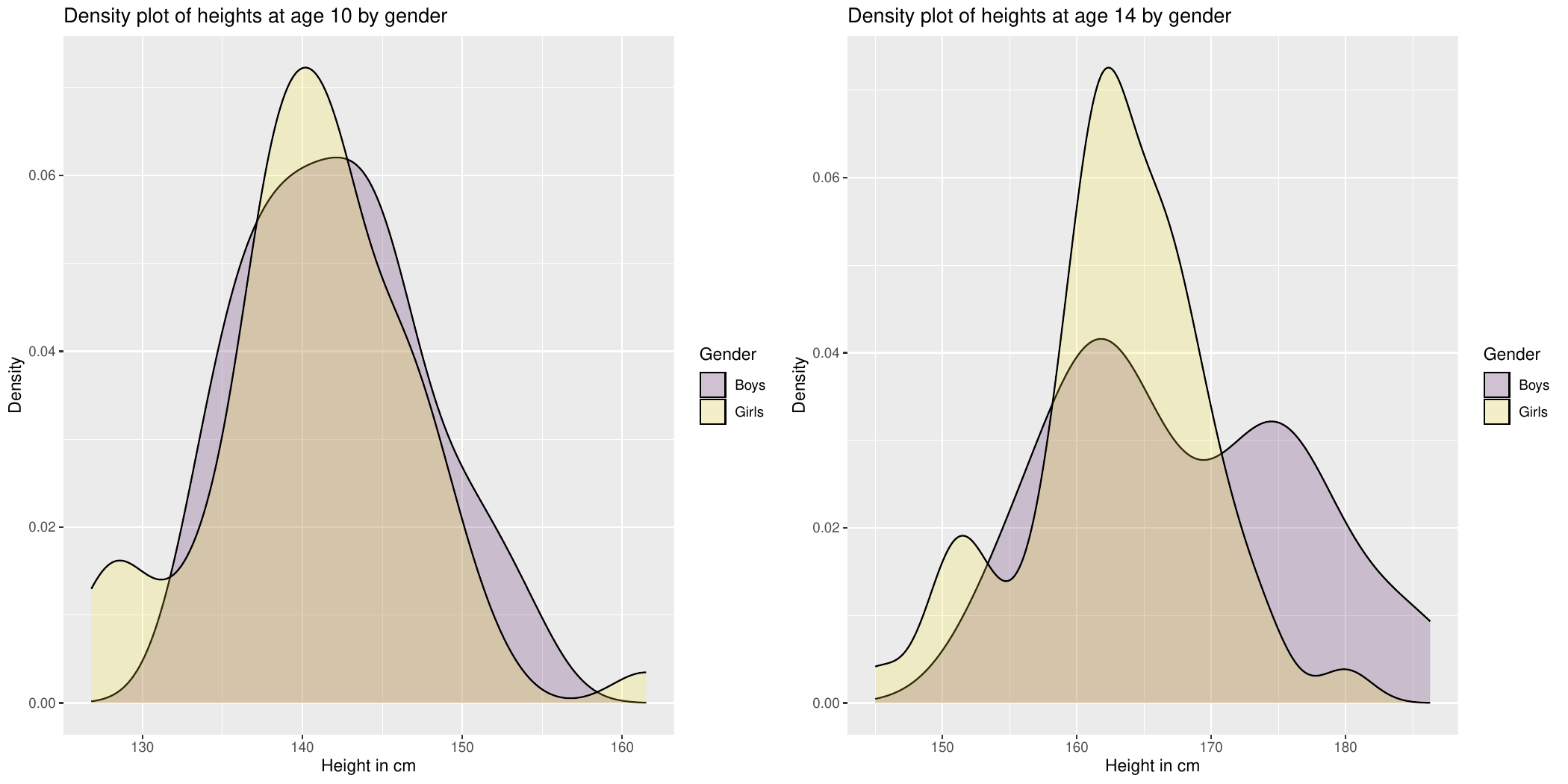}
    \caption{Density plots for the heights in cm of boys and girls at the age of 10 (left) and at the age of 14 (right).}
    \label{fig:height_densities}
\end{figure}

The Berkeley growth dataset consists of height measurements from 39 California boys and from 54 California girls in centimeters. The height of each child was followed and measured at multiple points over time, from the first year of their life until adulthood. In particular, for the first year, the children were measured every three months, then once a year until the age of 8, and finally every 6 months until the age of 18. The data were initially collected to understand the factors influencing human growth \citep{tuddenham1954physical}. The densities of the heights in centimeters for the two genders at the ages of 10 and 14 are presented in Figure \ref{fig:height_densities}. The plot indicates that boys are generally taller than girls, but still unclear if this difference is significant. The dataset is available in the R package fda \citep{Ramsey2023}. 

\begin{figure}[H]
    \centering
    \includegraphics[scale=0.45]{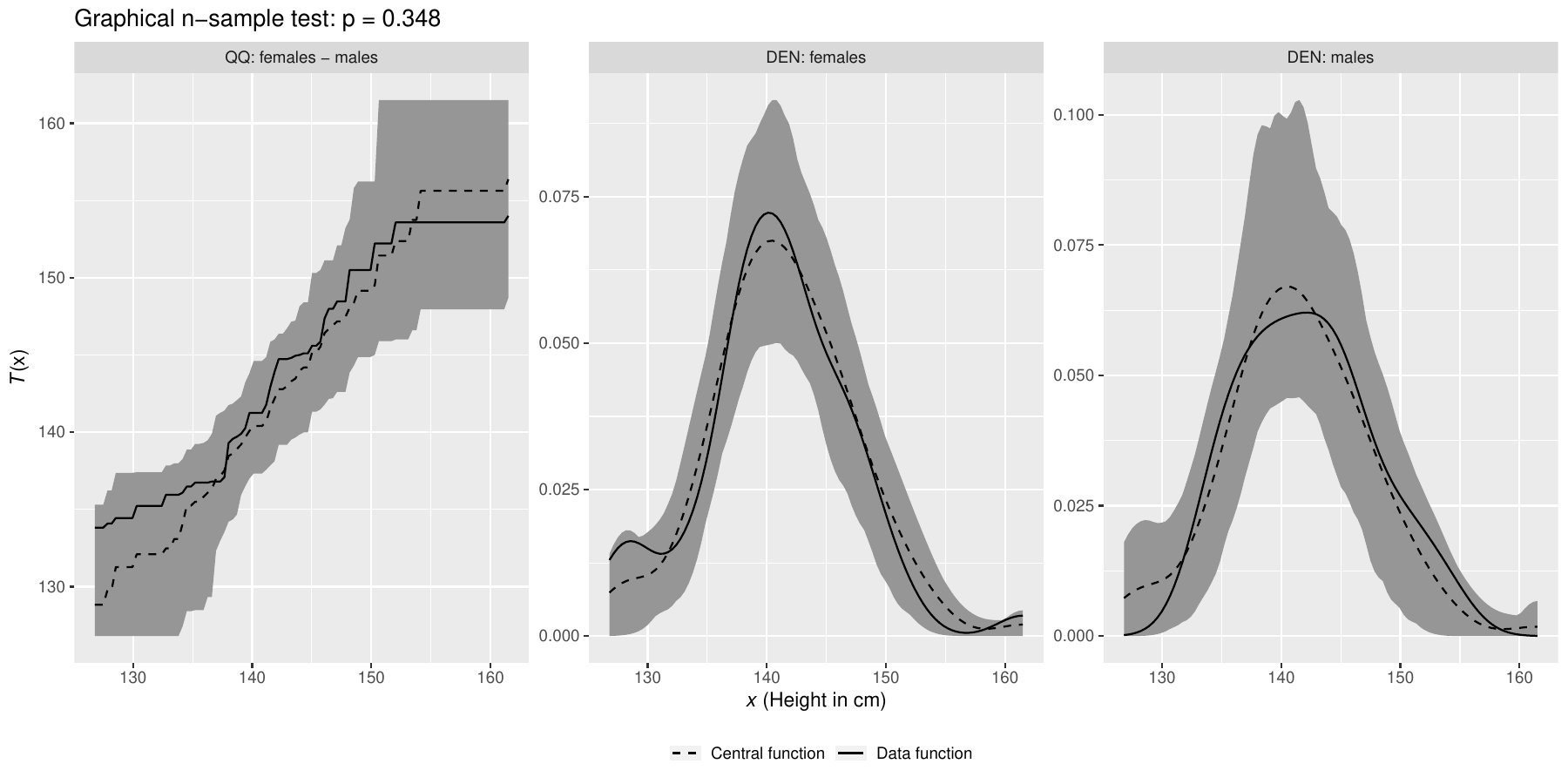}
    \caption{Graphical interpretation of the combined global envelope test for height distributions of boys and girls at the age of 10. The test was based on 1000 permutations.}
    \label{fig: heights10}
\end{figure}

Here, we are interested in investigating the hypothesis of equality between the height distributions of the two genders at different ages. For this purpose, we used the $\text{C}2$ test, i.e., a combined global envelope test with test statistics  $\mathbf{T}^{\text{qq}}$ and  $\mathbf{T}^{\text{den}}$. Regarding the height distributions of the two genders at the age of 10 the test suggests that there is not enough evidence to reject the null hypothesis of equality of the two distributions ($p = 0.348$). A graphical interpretation of the test result is presented in Figure \ref{fig: heights10}. The empirical test statistics are fully contained within the $95\%$ global envelopes indicating that the height distributions of boys and girls at the age of 10 are the same.

\begin{figure}[H]
    \centering
    \includegraphics[scale=0.45]{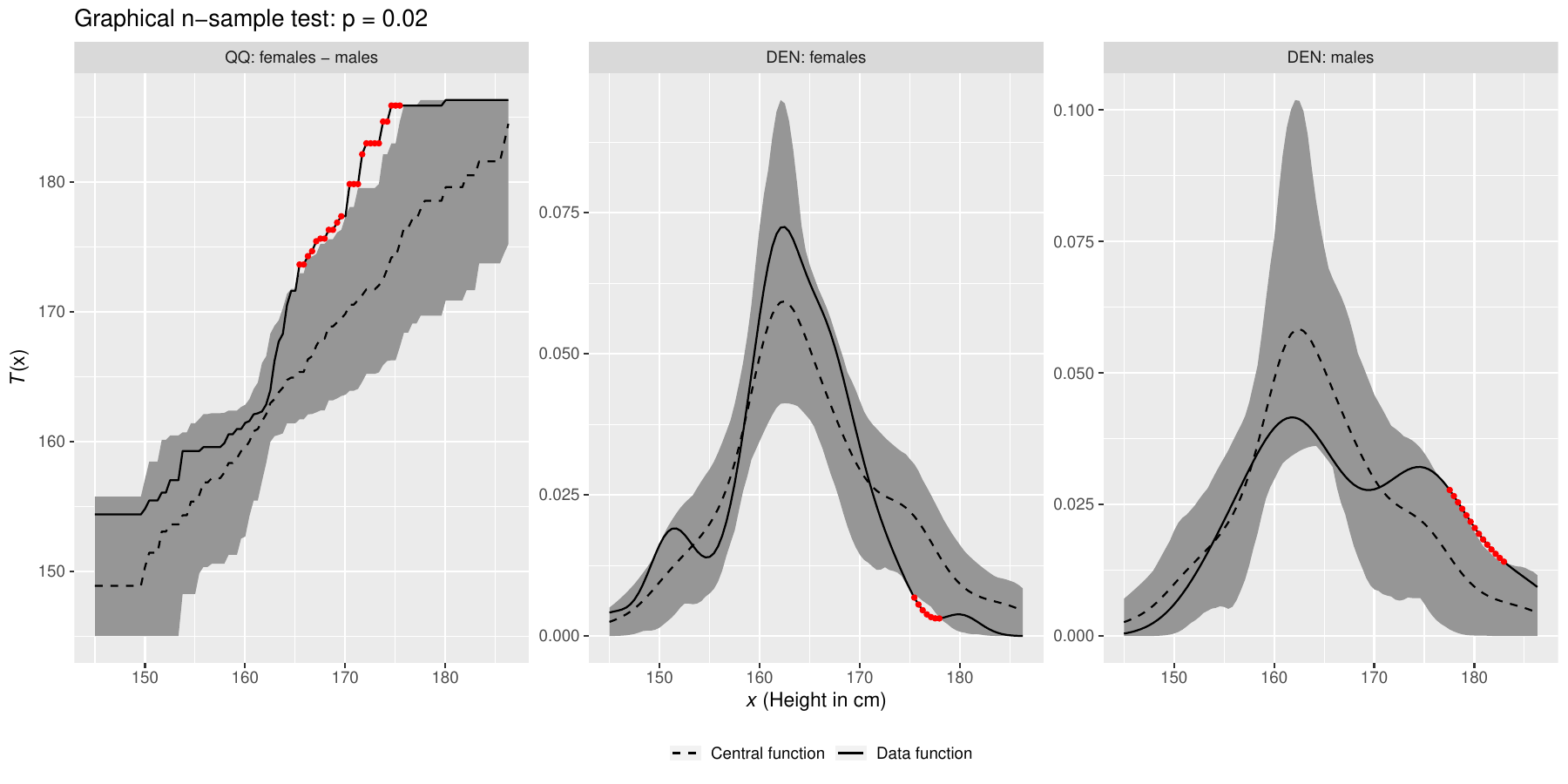}
    \caption{Graphical interpretation of the combined global envelope test for height distributions of boys and girls at the age of 14. The test was based on 1000 permutations.}
    \label{fig: heights14}
\end{figure}

Further, we applied the same test to compare the height distributions of the two genders at the age of 14. According to our results, the height distributions of the two genders are different ($p=0.02$, Figure \ref{fig: heights14}). The graphical interpretation of the test allows us to characterize this difference: There is a significantly higher proportion of boys with a height higher than 175 cm and a significantly lower proportion of girls with a height higher than 175 cm than what is expected under the null hypothesis. This is proven both by $\mathbf{T}^{\text{qq}}$ and  $\mathbf{T}^{\text{den}}$. This indicates that at the age of 14, there are more tall boys than tall girls.

\subsection{Iris data}
\begin{figure}[H]
    \centering
    \includegraphics[scale=0.5]{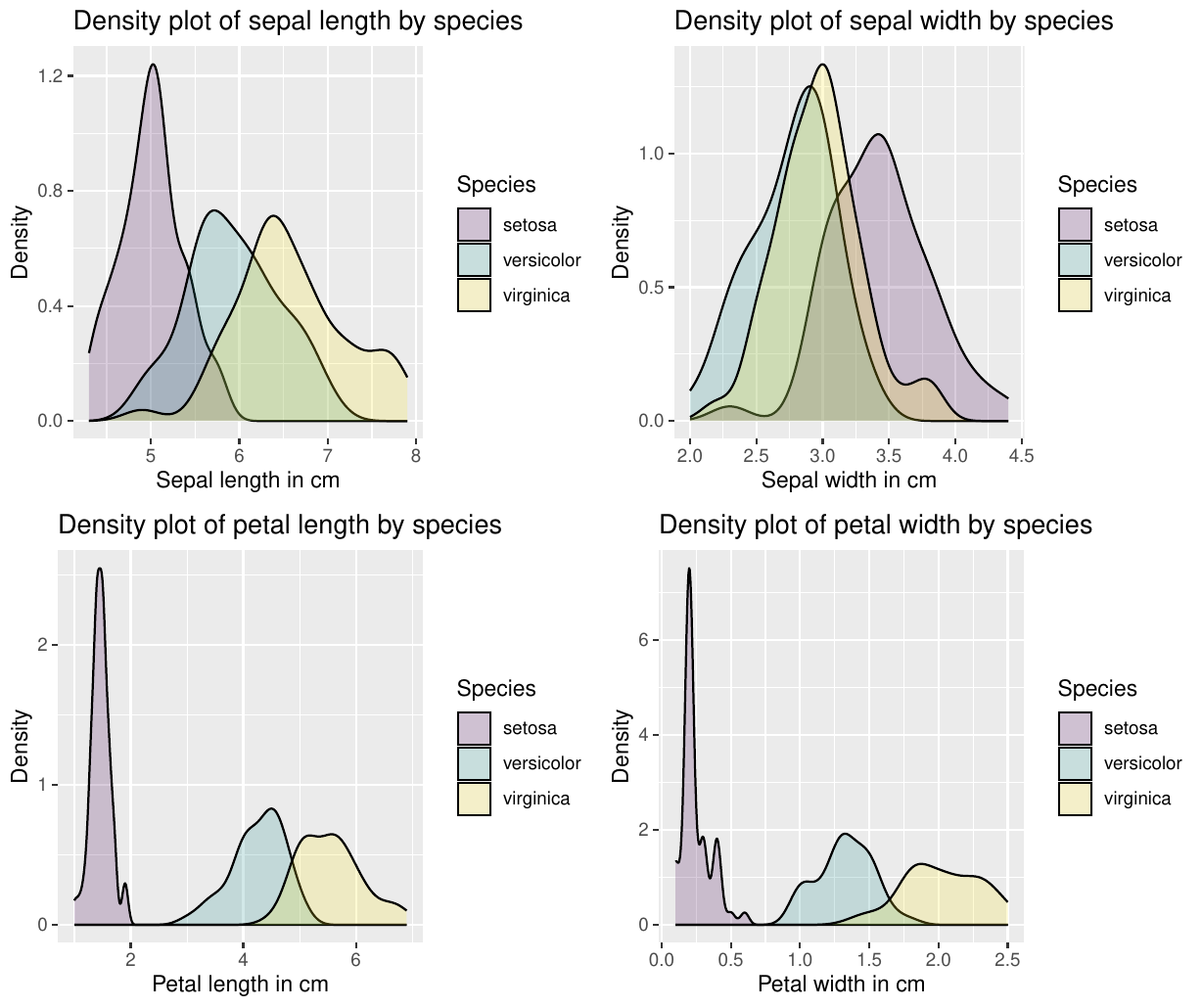}
    \caption{Density plots for the sepal length and width (top row) and petal lengths and width (bottom row) for the three iris flower species.}
    \label{fig:iris_densities}
\end{figure}

\begin{figure}[H]
    \centering
    \includegraphics[scale=0.5]{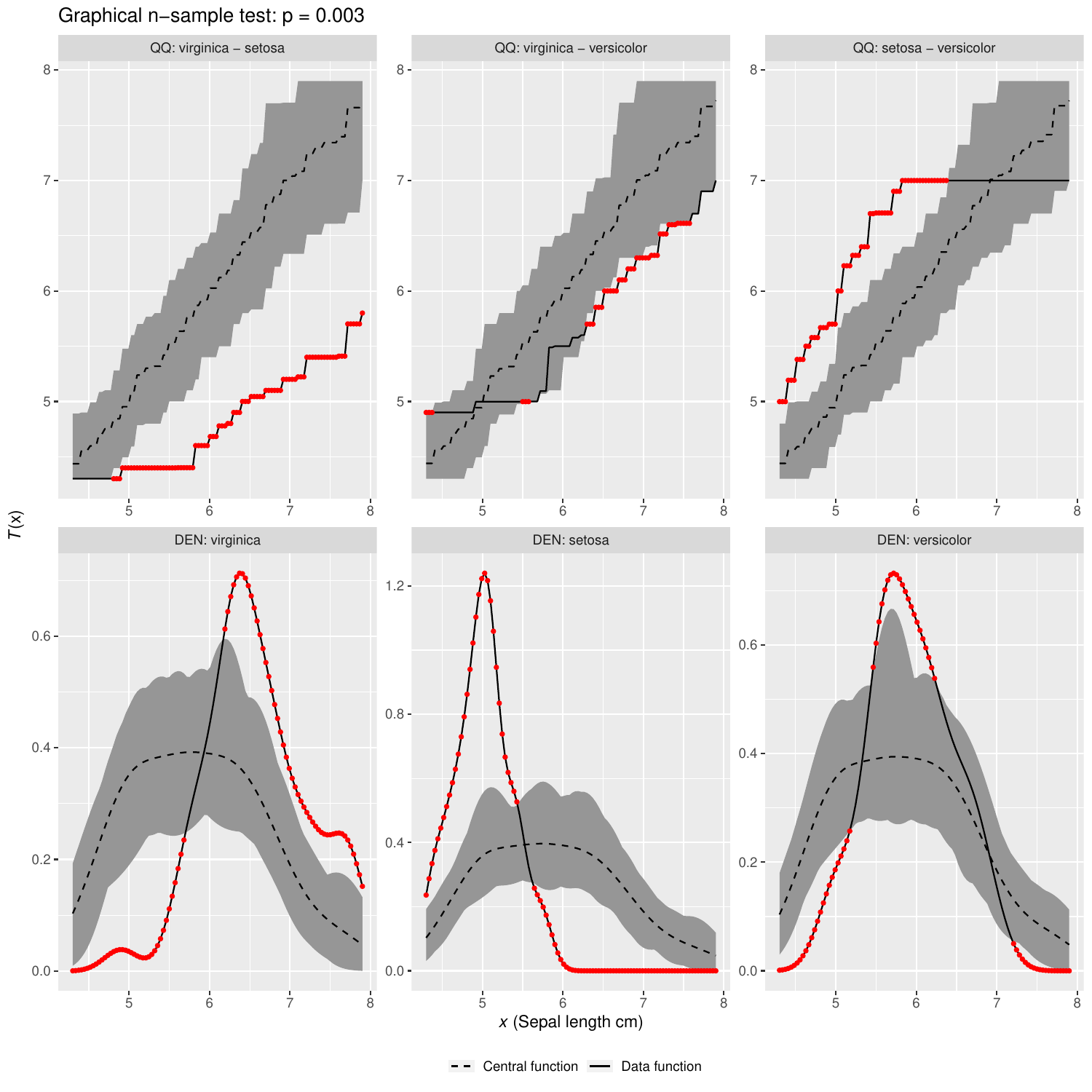}
    \caption{Graphical interpretation of the combined global envelope test for the sepal lengths distributions of three flower species (setosa, virginica and versicolor). The test was based on 1000 permutations.}
    \label{fig: iris}
\end{figure}

The iris dataset consists of 150 measurements from iris flowers growing together in the same colony. The data were collected by Dr. Edgar Anderson \citep{anderson1935irises} and were originally studied by the famous statistician Ronald Fisher \citep{fisher1936use} as a means of introducing discriminant analysis.  The dataset includes flower characteristics from 50 flowers from three flower species. Flower species under study are iris setosa, iris virginica, and iris versicolor. For each flower, measurements of the sepal length, sepal width, petal length, and petal width in centimeters are available.  The measurements from the iris setosa flowers are generally smaller, the measurements from the iris versicolor are intermediate in size, and the iris virginica are larger in size (see Figure \ref{fig:iris_densities}).  The dataset is publicly available as an R dataset \citep{R2023}. 
Here, we investigated distributional differences in the sepal length distributions between the three flower species (top left of Figure \ref{fig:iris_densities}). For this purpose, we used the $\text{C}2$ test with $n=3$, i.e., the combined global envelope test with test statistics  $\mathbf{T}^{\text{qq}}$ (see Equations \eqref{eq:Tqq_lk} and \eqref{eq: npairs}) and $\mathbf{T}^{\text{den}}$ (see Equations \eqref{eq:Tden} and \eqref{eq:nsampleT}) and $n=3$.
Then, a combined test (see Section \ref{sec: combining}) was constructed based on the test vector
\begin{equation*}
\mathbf{T}=(\mathbf{T}^{\text{qq}},\mathbf{T}^{\text{den}}).
\end{equation*}
The test rejects the null hypothesis of equality of the sepal length distributions of the three groups ($p < 0.001$, Figure \ref{fig: iris}). Moreover, the test provides a graphical interpretation of the result which can be used to characterize the distributional differences between each pair of groups (see Section \ref{sec:n-sample_tests}). All deviation patterns are similar to the one shown in Figures \ref{fig: graphical_mean} and \ref{fig: graphical_var}. Thus we can conclude that the differences are mostly in the mean and the variance.

\subsection{Log-returns for exchange courses}

Exchange rates are the rates for which a currency can be exchanged for another currency. Here, we considered the log returns of the EUR/USD and EUR/TRY courses between 2017 and 2019. This choice was made to avoid the highly volatile period during the Covid-19 pandemic. 
In this time-series example, the tests are not directly applicable as the assumption of iid samples is not valid due to the temporal dependence of the data. Therefore, we applied the test to the standardized residuals of a specific time series model.

We considered a standard generalized autoregression conditional heteroscedastic model (GARCH) of orders 1 and 1 for the variance \citep{bollerslev2008glossary}, and an autoregressive moving average model (ARMA) of order p and q model for the mean. That is, we assumed the following model for the time series of the log-returns $y_t$ at time t
\[y_t \sim N(\mu_t,\sigma^2_t)\]
where the conditional mean $\mu_t$ is modeled by an ARMA(p,q) model
\[
\mu_t = \sum_{i=1}^p \phi_i \mu_{t-p} + \sum_{i=1}^q \theta_i \epsilon_{t-p},
\]
where $\epsilon_k\sim N(0,\sigma^2_k)$ for $k = t -1 ,\ldots, t - p$, and the conditional variance is modeled as
\[\sigma^2_t = w + a_1\epsilon^2_{t-1}+b_1\sigma^2_{t-1}\]
with $w > 0, a_1, b_1 \geq0$ and $a_1 + b_1 <1$.  To fit the Garch model we used the rugarch package in R \citep{rugarch}.

\begin{figure}[H]
    \centering
    \includegraphics[scale=0.6]{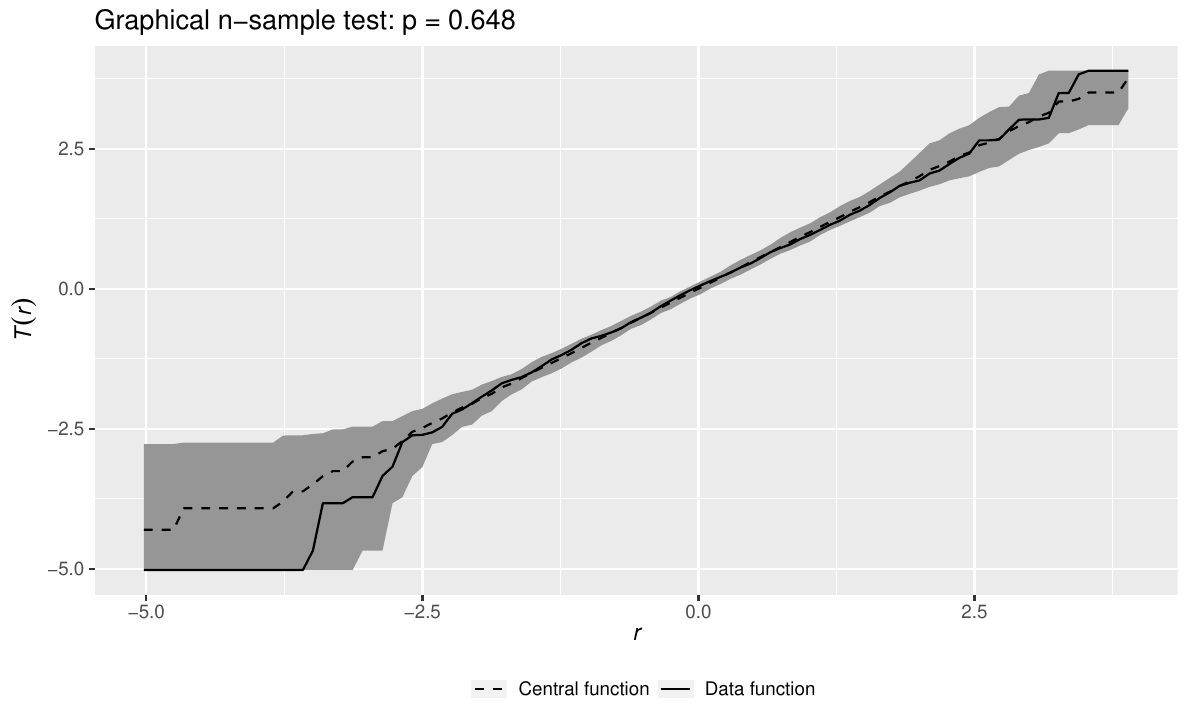}
    \caption{Comparison between the standardized residuals of a Garch(1,1) model with an ARMA(p,q) model for the conditional mean fitted to the log-returns of the EUR/USD course, with a sample of the same size from the N(0,1) distribution. The test was based on 1000 simulations. }
    \label{fig: euro_to_usd}
\end{figure}

According to the literature, if the model is sufficient in describing the data, the standardized residuals $z_t = \frac{\epsilon_t}{\sigma_t}$ should be an iid sequence from a white noise process, here the $N(0,1)$ distribution. To this end, we used the QQ test to compare the distribution of the standardized residuals of two Garch(1,1) models fitted to the log returns of the EUR/USD exchange courses against the distribution of a random sample of the same size from the $N(0,1)$ distribution. As seen from Figure \ref{fig: euro_to_usd}, the standardized residuals of the model are indistinguishable from iid realizations from a standard Normal distribution (p = 0.346) indicating that the model is adequate at modeling the log returns for the EUR/USD course.

\begin{figure}[H]
    \centering
    \includegraphics[scale=0.6]{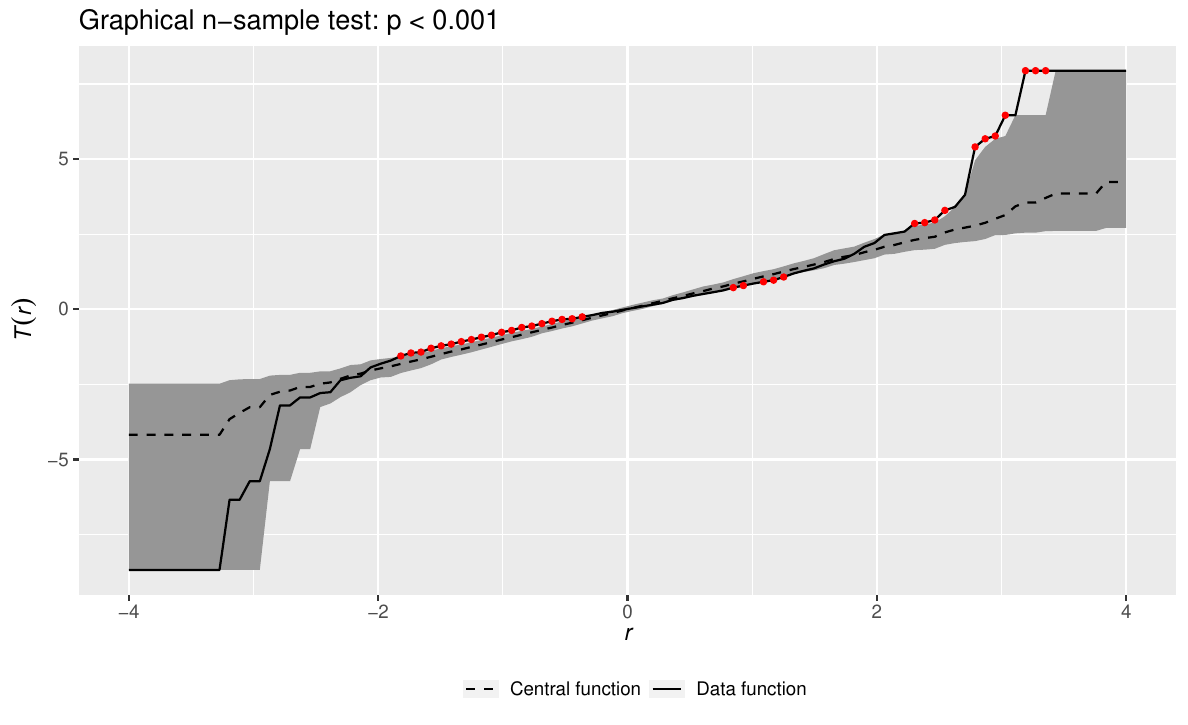}
    \caption{Comparison between the standardized residuals of a Garch(1,1) model with an ARMA(p,q) model for the conditional mean fitted to the log-returns of the EUR/TRY course and a sample of the same size from the $N(0,1)$ distribution. The test was based on 1000 permutations. }
    \label{fig: euro_to_try}
\end{figure}

Furthermore, we fitted the previously mentioned time series model to the log-returns of the EUR/TRY course. Figure \ref{fig: euro_to_try} compares the standardized residuals of the model against iid realizations from a standard Normal distribution. According to the results, the standardized residual distribution is different than the  standard Normal distribution $(p < 0.001)$, indicating that the model is inadequate at modeling the log returns for the EUR/TRY course. In particular, the standardized residuals have a distribution with heavier tails than the N(0,1) distribution. 

\section{Discussion}\label{sec:discussion}

In this paper, we proposed non-parametric, permutation-based tests for comparing the distribution of $n$ samples. The tests are based on the global envelope testing procedure which was introduced by \cite{MyllymakiEtal2017} to solve the multiple testing problem in spatial statistics. The tests not only provide a Monte Carlo $p$-value but also provide a graphical interpretation of the test result. This allows the user to investigate the reason for rejecting the null hypothesis and get useful insights into the data at hand. The test is global in the sense that the test is performed simultaneously for all discrete values of the discretization of the test statistic, as well as all test statistics (the case of combined tests) and all group comparisons (the case of $n$ samples). That is, given a prespecified global significance level $\alpha$, a global envelope test constructs an acceptance region by controlling the family-wise error rate. Recently, methods for constructing the acceptance region by controlling the false discovery rate instead of the family-wise error rate were proposed by \cite{mrkvivcka2023false}. It is useful for identifying all differences between the distributions since it is designed for local testing under the control of the expected number of false discoveries.

The proposed framework is generic and can be applied to any functional or multivariate test statistic $\mathbf{T}$ as the tests are based on ranks.
That is, the test achieves the correct nominal level independently of the distribution of the test statistics, as long as the test statistics can be strictly ordered; in the case of ties in the test statistics, the test is slightly conservative. These ties tend to be rather rare, but they are possible in a practical situation. The only assumption needed is the exchangeability of the test statistics under the method used to simulate data under the null model. Here, we used a simple permutation scheme to simulate new data; that is, the observations of the variable of interest were permuted between the samples. The exchangeability assumption is satisfied for this permutation strategy. 

We conducted a simulation study where we investigated the power of the proposed testing procedure with different test statistics capturing different aspects of the underlying distributions. We considered five test statistics as well as some combinations of them. The tests were applied to the two-sample cases under different simulated scenarios. In each scenario, a different distributional difference was investigated. Moreover, in each simulated experiment, we provided guidelines on how one should interpret the test results for the different proposed test statistics.

The statistical power of the global tests was compared to the power of the asymptotic Kolmogorov-Smirnov test, as this test also provides a graphical interpretation of the test result. Overall, the proposed tests outperformed the classical Kolmogorov-Smirnov test as they achieved higher statistical power in all studied settings. Even though the combined tests performed quite well in all simulated experiments, they were slightly outperformed by
the best global tests based on single statistics. However, usually in real applications, distributional differences are a priori non-characterized, and therefore, using a combined test is suggested. The combined test also brings a wider graphical interpretation.

Last but not least, we demonstrated how the proposed tests can be applied to real data examples. For this purpose, we considered three data examples, the Berkeley growth dataset, the iris dataset, and an exchange course example. For the Berkeley growth data, we applied a combined global envelope test and compared the height distributions between the boys and girls at different ages. For the iris data, we applied a combined global envelope test and investigated whether there are distributional differences between the sepal length distributions of three flower species. In the exchange rates example, we investigated the distributions of the standardized residuals from a time series conditional heteroscedastic model fitted to the log returns of specific courses. In the future, we aim to investigate whether the proposed tests can be generalized to the case in which nuisance covariates are present with influence on the distribution of the variable under study. This problem is already solved for the quantile regression test statistic using the framework suggested by \cite{mrkvivcka2023global}, but remains unsolved for a generic test statistic. The suggested procedure utilizes more sophisticated permutation strategies to remove the nuisance effects from the response distribution.

\section*{Acknowledgements}

This study has been done under the Research Council of Finland's flagship ecosystem for Forest-Human-Machine Interplay---Building Resilience, Redefining Value Networks and Enabling Meaningful Experiences (UNITE) (Grant number 357909). The authors also acknowledge funding from the Swedish Research Council (VR 2018-03986).
The authors wish to acknowledge CSC – IT Center for Science, Finland, for computational resources. The authors also thank Aila S\"arkk\"a for valuable feedback and suggestions.

\bibliographystyle{chicago}

\appendix

\section*{Appendix A: Type I errors }
\label{sec: appendixA}

\begin{figure}[H]
    \centering
    \includegraphics[scale=0.7]{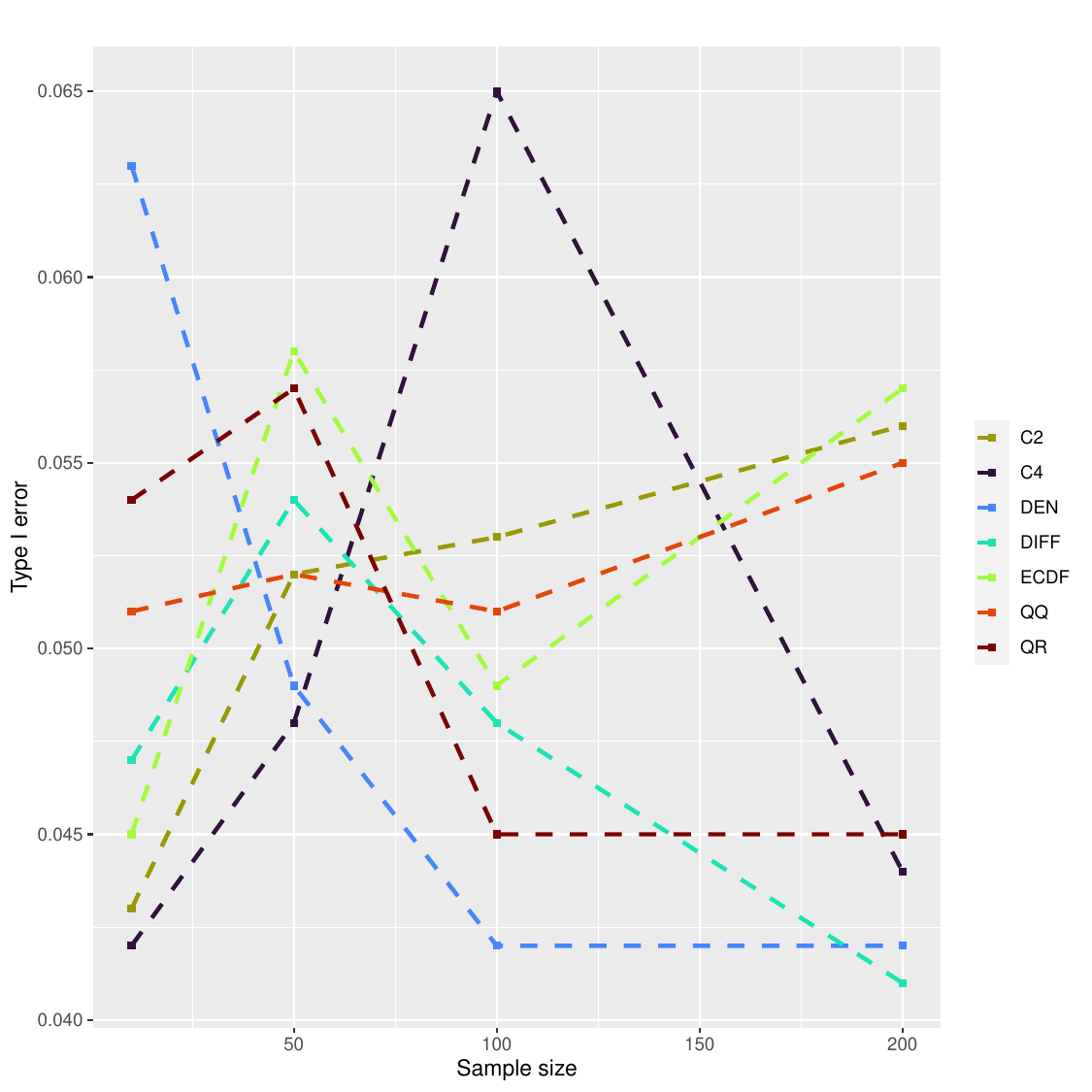}
    \caption{Type I errors of the tests among 1000 simulated standard normal samples of different sizes (x-axis) for the different tests of Table \ref{table: tests} (colors).}
    \label{fig: TypeIerror}
\end{figure}
\end{document}